\documentclass[a4paper,12pt]{article}
\usepackage{amsmath,amssymb}
\usepackage{hyperref}
\usepackage{graphicx}
\usepackage{mathrsfs}
\usepackage{float}
\usepackage{ulem}
\usepackage{braket}

\hypersetup{
    colorlinks,%
    citecolor=blue,%
    filecolor=blue,%
    linkcolor=blue,%
   urlcolor=blue,
   linktoc=page
}
\usepackage{chngcntr}
\counterwithin{equation}{section}
\usepackage{cite}

\newcommand{\be}{\begin{equation}}
\newcommand{\ee}{\end{equation}}
\newcommand{\f}{\frac}
\newcommand{\s}{\sqrt}

\newcommand{\bea}{\begin{equation}\begin{aligned}}
\newcommand{\eea}{\end{aligned}\end{equation}}
\newcommand{\ba}{\begin{align}}
\newcommand{\ea}{\end{align}}

\newcommand{\la}{\langle}
\newcommand{\ra}{\rangle}
\newcommand{\beq}{\begin{equation}}
\newcommand{\eeq}{\end{equation}}

\begin{document}

\begin{titlepage}

\vspace{.4cm}
\begin{center}
\noindent{\Large \textbf{Chaos and relative entropy}}\\
\vspace{1cm}
Yuya O. Nakagawa$^{a,}$\footnote{y-nakagawa@15.alumni.u-tokyo.ac.jp}, $\quad$
G\'abor S\'arosi$^{b,c,}$\footnote{sarosi@sas.upenn.edu},
$\quad$
 Tomonori Ugajin$^{d,}$\footnote{tomonori.ugajin@oist.jp}

\vspace{.5cm}
 {\it
 $^{a}$ Institute for Solid State Physics, the University of Tokyo,\\ Kashiwa, Chiba 277-8581, Japan\\
\vspace{0.2cm}
 }
 \vspace{.5cm}
  {\it
  $^{b}$David Rittenhouse Laboratory, University of Pennsylvania,\\
  Philadelphia, PA 19104, USA\\
\vspace{0.2cm}
}
\vspace{.5cm}
 {\it
 $^{c}$Theoretische Natuurkunde, Vrije Universiteit Brussels,  \\
Pleinlaan 2, Brussels, B-1050, Belgium\\
\vspace{0.2cm}
 }
\vspace{.5cm}
  {\it
 $^{d}$Okinawa Institute of Science and Technology, \\
Tancha, Kunikashira gun,  Onna son, Okinawa 1919-1 \\
\vspace{0.2cm}
 }
\end{center}


\begin{abstract}
One characterization of a chaotic 
system is the quick delocalization of quantum information (fast scrambling). One therefore expects that in such a system a state quickly becomes locally indistinguishable from its perturbations. In this paper we study the time dependence of the relative entropy between the reduced density matrices of the thermofield double state and its perturbations in two dimensional conformal field theories. We show that in a CFT with a gravity dual, this relative entropy exponentially decays until the scrambling time. This decay is not uniform. We argue that the early time exponent is universal while the late time exponent is sensitive to the butterfly effect.
This large $c$ answer breaks down at the scrambling time, therefore we also study the relative entropy in a class of spin chain models numerically. We find a similar universal exponential decay at early times, while at later times we observe that the relative entropy has large revivals in integrable models, whereas there are no revivals in non-integrable models.  
\end{abstract}

\end{titlepage}

\tableofcontents

\hspace{0.3cm}  


\section{Introduction}


There are many different notions of chaos with a somewhat limited understanding of the relation between them. Some of these are
\begin{enumerate}
\item For classical systems, under chaos we usually mean some notion of ergodicity of the dynamics. One signal of ergodicity is chaotic mixing of phase space trajectories, which is related to the heavy dependence of them on small perturbations of the initial conditions \cite{ott2002chaos}. This is called the butterfly effect and is characterized by a so called Lyapunov exponent $\lambda_L$, setting the speed at which nearby trajectories diverge at early times.
\item For quantum systems, there is a notion of thermalization, which means that simple (time ordered) correlators relax to their thermal values if the system is started from some non-equilibrium state. This is an early time effect that happens at times of order $\beta=T^{-1}$, where $T$ is the temperature.
\item \label{pt:Lyapunov}For quantum systems with a classical limit controlled by some tuneable parameter $\chi$, the classical butterfly effect is related to the exponentially decaying behaviour of out of time order correlators (OTOC) \cite{larkin1969quasiclassical} at times smaller than the so called Ehrenfest time $t_E=\frac{1}{\lambda_L}\log\frac{1}{\chi}$ \cite{combescure1997semiclassical}. For general quantum systems, the OTOCs can still have Lyapunov type behaviour and the rate of their decay is bounded as $\lambda_L \leq 2\pi T/\hbar$ because of causality and unitarity constraints \cite{Maldacena:2015waa}. The Lyapunov behaviour happens at intermediate time scales, which are much longer than the thermalization time. Note that $\chi$ might be $\hbar$ in which case the bound is trivial in the classical limit, but this is not necessary. For example, in AdS/CFT one has $\chi =N^{-2}$.
\item Another, intrinsically quantum notion of chaos is the randomness of the energy spectrum, more precisely, the level spacing statistics, which is said to be chaotic if it agrees with that of random matrix theory \cite{mehta2004random}, in particular when nearby energy levels repel each other \cite{dyson1962statistical1,dyson1962statistical2}. Since this phenomenon is sensitive to the discreteness of the spectrum, it is associated to effects at very late times, exponential in the entropy.\footnote{See also \cite{Cotler:2016fpe} in the context of black holes.}
\item For quantum systems with some locality structure, there are notions like the eigenstate thermalization hypothesis, which is the statement that energy eigenstates appear thermal when probed by sufficiently simple and local probes \cite{deutsch1991quantum,srednicki1994chaos,rigol2008thermalization}. 
\item \label{pt:scr} Again for quantum systems with a notion of locality, there is the phenomenon of scrambling of localized quantum information \cite{Sekino:2008he}. At the intuitive level, this is related to classical notions of ergodicity and divergence of trajectories, as both these measure how mixing the dynamics is, and how much it forgets about initial conditions. There is both the question of the speed of scrambling (measured by some scrambling time) and how effectively does it happen, i.e. how scrambled localized information can get. There are many quantities sensitive to this type of physics. For example, the previously mentioned OTOCs are also sensitive to this at late times, because they can be regarded as simple measures of operator growth in the sense of Lieb-Robinson bounds \cite{lieb1972finite,nachtergaele2006propagation,bravyi2006lieb,Roberts:2016wdl}.\footnote{While OTOCs are always sensitive to scrambling, the butterfly effect only makes sense if there is a classical limit of the system. It is not entirely clear what is the precise connection between these two types of physics, in particular notions like the scrambling time (the time when initally localized information gets maximally scrambled) and the Ehrenfest time (the time when a wavepacket spreads so much that the classical approximation breaks down). In holographic systems, the two timescales are the same basically because the parameter controlling the classical limit is also related to the number of degrees of freedom. In a generic system with a classical limit such relation does not necessarily exist. } Beyond this, quantum information theoretic quantities, like the trace distance \cite{Lashkari:2011yi}, the mutual information or the tripartite information \cite{Hosur:2015ylk} are also sensitive to scrambling.
\end{enumerate} 

The primary aim of this paper is to add another quantity to points \ref{pt:Lyapunov} and \ref{pt:scr}, which is sensitive to scrambling and possibly the Lyapunov behaviour, namely the relative entropy
 of reduced density matrices associated with a local subregion.  The relative entropy $S(\rho||\sigma)$ measures the distinguishability of two density matrices $\rho$ and $\sigma$, and is defined by\footnote{The relative entropy has been used efficiently in the recent quantum information theoretic approach to some fundamental questions in quantum field theory \cite{Faulkner:2016mzt,Casini:2016udt,Casini:2017vbe,Casini:2017roe,Balakrishnan:2017bjg} and quantum gravity \cite{Faulkner:2013ica,Lashkari:2014kda,Jafferis:2015del,Dong:2016eik,Lashkari:2016idm, Faulkner:2017vdd,Faulkner:2017tkh}.} 
\be  
S(\rho||\sigma) = {\rm tr} \rho \log \rho -  {\rm tr}\rho \log\sigma  . 
\ee
Note that $S(\rho||\sigma) =0$ implies $\rho=\sigma$.  When a system scrambles, i.e. quantum information becomes quickly delocalized, the reduced density matrices $\rho_{\phi}, \rho_{\psi}$ of two states  $|\phi \ra, | \psi \ra$ of similar energy become hardly distinguishable after the scrambling time without having access to a large fraction of all the degrees of freedom. Based on this, we expect the relative entropy on a spatial subsystem to show a decaying behaviour, with the rate of the decay quantifying the speed of scrambling, while the late time value, after the decay ends, quantifying how scrambled the initially localized information can get. In a chaotic system, we therefore expect this late time value to be small.\footnote{Considering the relative entropy as an indicator of scrambling is very similar to using the trace distance, as done in \cite{Lashkari:2011yi}. In fact, the relative entropy is a more refined probe because of Pinsker's inequality \cite{ohya2004quantum}.}

To sharpen this intuition, we could think about scrambling as the question of how effectively can we recover information from a state after the application of a quantum channel $\mathcal{N}_t$ which consists of time evolution followed by a partial trace over some spatial region $B$
\be
\rho \mapsto \mathcal{N}_t(\rho)=\text{Tr}_B\big( e^{-iHt}\rho e^{iHt}\big).
\ee
For such noninvertible channels, there exist approximate recovery maps. However, the possible effectiveness of such recovery channels is bounded by the relative entropy \cite{junge2015universal}\footnote{See also \cite{Cotler:2017erl} for a use of this bound for bulk reconstruction.}
\be
\label{eq:recoverybound}
S(\rho||\sigma)-S(\mathcal{N}_t(\rho)||\mathcal{N}_t(\sigma)) \geq -2 \log F(\rho,\mathcal{R}_{\sigma,\mathcal{N}_t}(\rho)),
\ee
where $\rho$ and $\sigma$ are any two states and $\mathcal{R}_{\sigma,\mathcal{N}_t}$ is a particular approximate recovery channel which can recover the state $\sigma$, while $F$ is the fidelity. In this sense, the time dependent relative entropy $S(\mathcal{N}_t(\rho)||\mathcal{N}_t(\sigma))$ tells us how quickly approximate recovery from this channel can fail.

\begin{figure}[h!]
\centering
\includegraphics[width=0.55\textwidth]{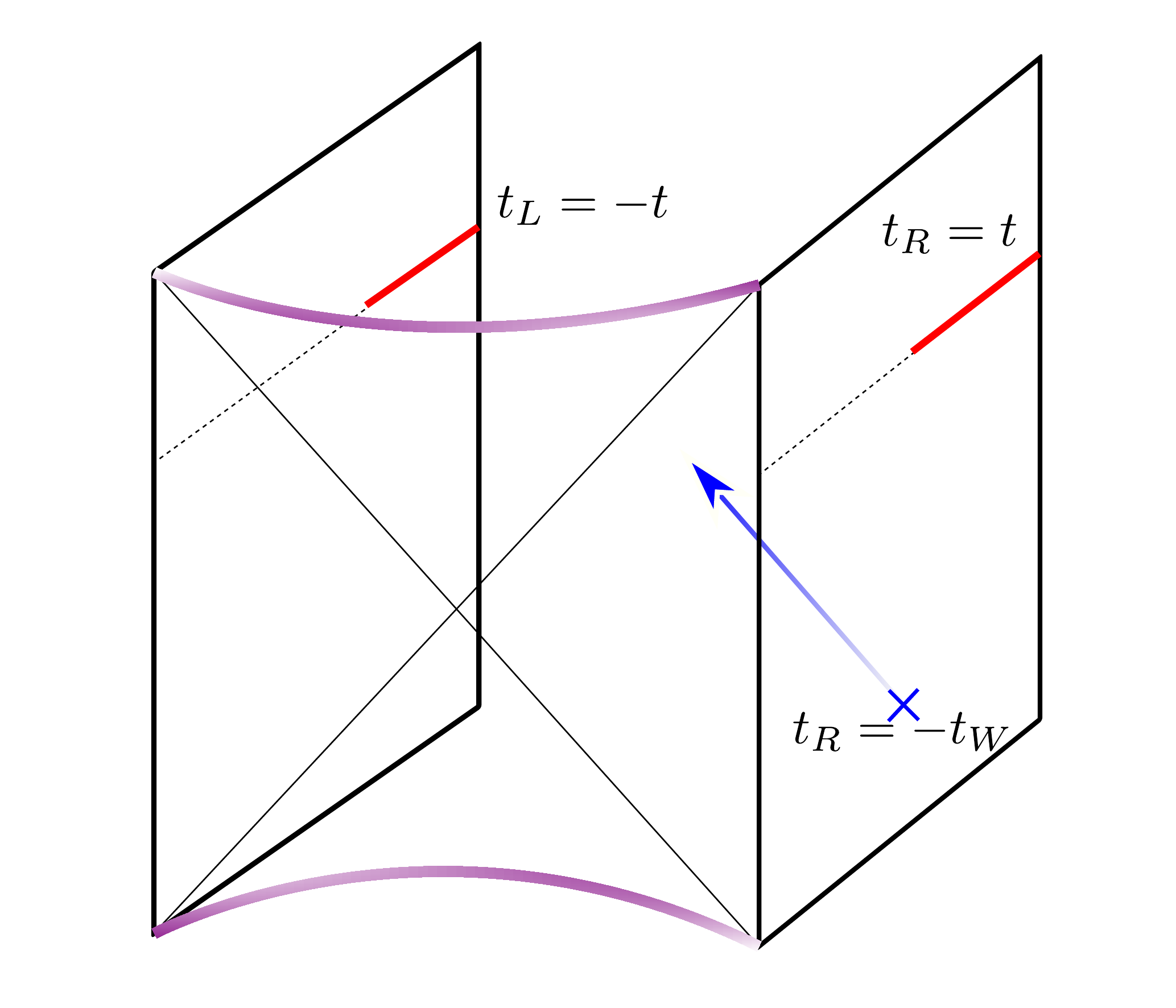}
\caption{We consider a setup in which the eternal black hole is perturbed by an operator insertion at early times (blue cross). We calculate the relative entropy of the subsystem drawn with red between the TFD state and its perturbation.}
\label{fig:0}
\end{figure}

The AdS/CFT correspondence \cite{Maldacena:1997re} gives an excellent tool to analytically study quantum chaos in strongly coupled systems. A particularly useful setup is the Shenker Stanford process \cite{Shenker:2013pqa}, in which one perturbs a thermofield double (TFD) state, which is holographically dual to a two sided eternal black hole, by injecting energy on one side.  In the dual gravity picture this amounts to sending a shock wave into the black hole. This process was argued to be chaotic, in particular, time evolution of the mutual information was calculated in \cite{Shenker:2013pqa}. In the present paper, we will be concerned with the relative entropy between the TFD state and its perturbation with the shockwave, both in the case of translational invariant and localized shocks. We take the spatial subsystem to be the union of the half line on both boundaries, see Fig \ref{fig:0}.

We will calculate this relative entropy in a holographic two dimensional conformal field theory (CFT) with large central charge and show that it indeed diagnoses scrambling.  The result is that the relative entropy is initially proportional to the central charge of the CFT, but it decays exponentially in time. Assuming that the subsystem is large enough,\footnote{When the subsystem is smaller than the scrambling time, its size gives the relevant timescale when the relative entropy drops to order one.} there are two different exponential behaviours. Initially, the decay goes as $\exp(-\frac{2\pi}{\beta} t)$ until times $\beta \ll t\sim \beta \log E$, where $E$ is roughly the total energy of the perturbation. In our setup, this energy will be large, but order $c^0$ in terms of the central charge $c$. We will argue that this decay rate is universal as it comes from the modular Hamiltonian piece. After this, the decay crosses over to $\exp(-\frac{4\pi}{\beta} t)$. We will argue that in a generic CFT, the rate of this second decay is related to the behavior of out of time order (OTO) correlators in the Lyapunov regime, so that it is sensitive to chaos. This argument comes from doing the calculation using the replica trick combined with large $c$ vacuum block techniques \cite{Fitzpatrick:2014vua}, besides directly applying the Ryu-Takayanagi formula \cite{Ryu:2006bv}. We will see that Lyapunov regime shows up in the replica correlators. This second decay continues until the relative entropy becomes of order one at the scrambling time $t\sim \beta \log c$, at which point quantum corrections to the Ryu-Takayanagi formula start to matter and we can no longer trust the result. We note that the exponential decay in time is something new compared to the typical linear or logarithmic time dependence of the entanglement entropy or the mutual information \cite{Calabrese:2005in, Calabrese:2007mtj, Hartman:2013qma,Caputa:2014eta,Caputa:2015waa}. 

Another interesting feature is the dependence on the time when the shockwave is inserted. The sooner the shockwave is inserted, the larger the perturbation to the TFD state is on the $t=0$ slice. In fact, a shockwave entering at time $t=-t_W$ results in a relative entropy proportional to $e^{\frac{2\pi}{\beta} t_W}$.\footnote{Note that the combined dependence on $t$, describing the location of the time slice, and $t_W$ describing the insertion of the shock is nontrivial, because the TFD state is not invariant under time translations.} This quantifies how far we end up with from the TFD state as a result of such an earlier perturbation, which has a similar flavour as the butterfly effect. We will argue however that this kind of dependence is universal in conformal field theories, so it is not directly related to the Lyapunov exponent. It would be very interesting though if this universal growth could be used to understand the chaos bound of \cite{Maldacena:2015waa} from an information theoretic point of view. We will show that this relative entropy bounds out of time order correlators, though unfortunately we were not able to relate to their rate of change.

In addition to the holographic results, we perform numerical calculation of this relative entropy in a spin chain model. The main observation is that after the decay in $t$ stops, the relative entropy stays small for a chaotic system, while has revivals comparable with the initial value for integrable systems.\footnote{Similar revivals for single sided global quenches in rational CFTs were studied in \cite{Cardy:2014rqa}. Also, time evolutions in the process and its recurrence was studied in \cite{Mandal:2016cdw}.}
   In this regard, it behaves similarly to the mutual information or the tripartite information. In addition to this, we will observe that the early time decay is exponential both for the chaotic and integrable cases. The exponent is proportional to the temperature similarly to the CFT case.

The organization of this paper is as follows. In section \ref{section:CFTpart}, we explain our setup in the setting of two dimensional conformal field theories. We show using the replica trick that the relative entropy is determined by how the replica correlators analytically continue in the replica index in their Regge-limit. Then, we obtain a concrete formula by approximating these correlators with the large $c$ vacuum Virasoro block. We spend section \ref{sec:formdisc} explaining the features of this formula and making some comments about the expected time dependence for non-holographic large $c$ theories via a possible connection to the Maldacena-Shenker-Stanford (MSS) chaos bound. Sections \ref{sec:hologlobal} and \ref{sec:hololocal} are devoted to calculations of the relative entropy using the Ryu-Takayanagi formula, for translationally invariant and local perturbations respectively. In section \ref{sec:spinchains} we present our numerical results for the spin chain model. We have appendix \ref{app:outofcausal} complementing some calculations in sec. \ref{section:CFTpart}. In appendix \ref{sec:distinctstates} we describe a generalization of our relative entropy to the case when both states are deformations of the thermofield double, while in appendix \ref{app:disjoint} we generalize the holographic calculations of sec. \ref{sec:hologlobal} to the case when the subsystem has finite size.

\section{Localized perturbations}

In this section, we explain our setup in the CFT, and how to calculate the relative entropy of interest using correlation functions. These correlation functions have several distinct OPE-like limits, depending on the causality relation between the subsystem and the local perturbations. The discussion of this is very much similar to those found in \cite{Caputa:2014eta,Caputa:2015waa}, where the time evolution of entanglement entropy for local quenches on thermal backgrounds were studied. 

We will eventually focus on the relative entropy between a thermofield double state and its perturbations, where we know the exact expression of the modular Hamiltonian.
The relative entropy consists of the modular Hamiltonian part as well as the entanglement 
entropy part. We first explain the way to evaluate the modular Hamiltonian part with
again paying attention to the causality of the set up. 

The entanglement entropy part can be evaluated by a four point function involving twist operators in the cyclic orbifold of the original CFT. In a CFT with a gravity dual, this four point function can be well approximated by the vacuum Virasoro conformal block and the result agrees with the holographic one given by the Ryu-Takayanagi formula. In the next section we will use the four point function expression to discuss a possible connection between the time dependence of the relative entropy and the MSS chaos bound \cite{Maldacena:2015waa}.

\label{section:CFTpart}

\subsection{General replica setup}

We will consider a two dimensional conformal field theory (CFT) and the thermofield double state
\beq
|TFD\rangle = \frac{1}{\sqrt{Z}} \sum_n e^{-\beta E_n/2}|n\rangle_L |n\rangle_R \in CFT_L \otimes CFT_R.\label{eq:TFDs}
\eeq
One can create this state by cutting up in half the Euclidean path integral on the cylinder which calculates the thermal partition function. The two lines of the cut correspond to the left and the right copy of the CFT. When the CFT has a gravitational dual, the $|TFD\rangle$ state is dual to the two-sided AdS-Schwarzschild black hole, connecting the two boundaries. The $|TFD\rangle$ state is a special model of a non-equilibrium quench state with respect to the time evolution generated by $H_L+H_R$, where $H$ is the Hamiltonian of the single CFT \cite{Hartman:2013qma}.

\begin{figure}[h!]
\centering
\includegraphics[width=0.45\textwidth]{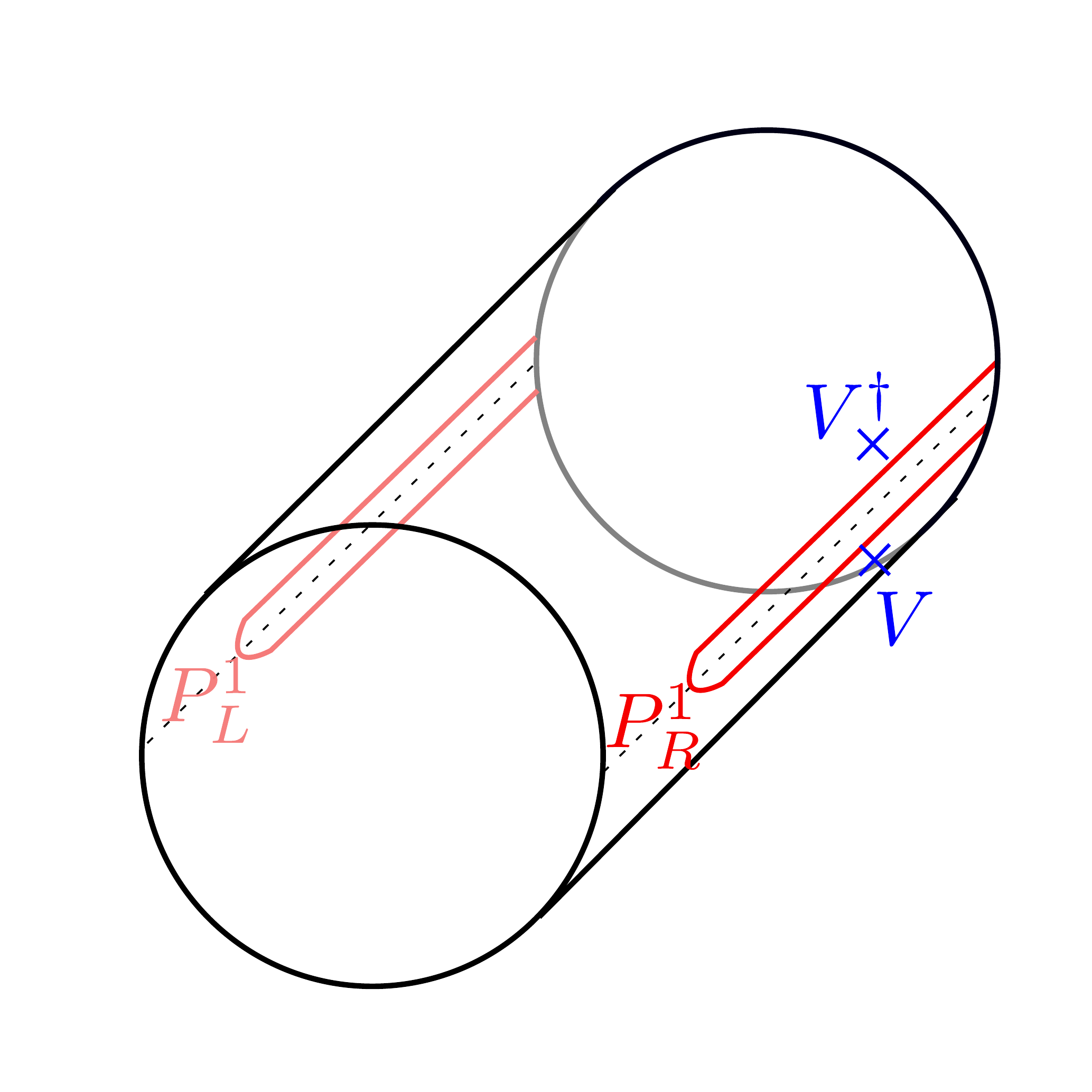}
\includegraphics[width=0.45\textwidth]{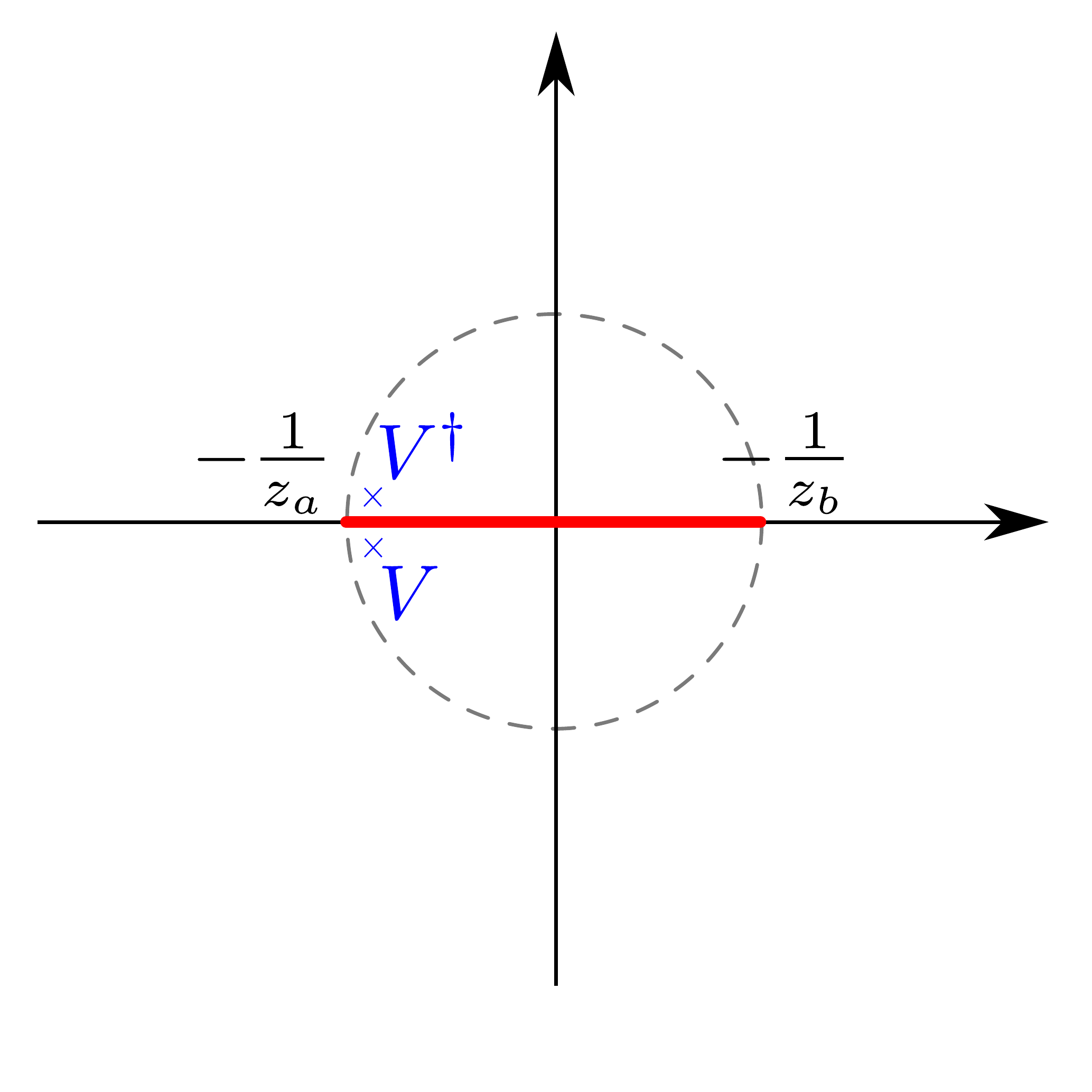}
\caption{Euclidean path integral creating the perturbed state. Left: drawn on the cylinder, red line is the cut along the subsystem, blue crosses are operator insertions. Right: drawn when mapped to the plane. Red line is the cut, blue crosses are operator insertions. Note that we have applied a global conformal map $z\mapsto -1/z$ on the figure compared to the text to make the location of the cut more illustrative.}
\label{fig:setup}
\end{figure}

We will be interested in local perturbations of the $|TFD\rangle$ state
\beq
V(x,t_p-i\epsilon)|TFD\rangle, \;\;\; W(x,t_p-i\epsilon)|TFD\rangle, \label{eq:TFDp}
\eeq
where $V$ and $W$ are local primary operators of the CFT. The role of the Euclidean time shift $\epsilon$ is to regulate the energy of these states and to make them normalizable. We will reduce these states to a subsystem which consists of a half line on both CFTs. The reduced density matrix is calculated as a path integral on the cylinder of circumference $\beta$ with cuts corresponding to the in and out indices of the matrix, see Fig. \ref{fig:setup}. The cuts are running from 
\beq
P^1_R:(x,t_E)=(0,it) \;\text{ to }\; P^\infty_R:(x,t_E)=(\infty,it)
\eeq
and from 
\beq
P^1_L:(x,t_E)=(0,-it+\beta/2) \;\text{ to }\; P^\infty_L(x,t_E)=(\infty,-it+\beta/2),
\eeq
 and operator insertions 
 \beq
 V \text{ at } (w_1,\bar w_1)=(x-t_p+i\epsilon,x+t_p-i\epsilon) \text{ and } V^\dagger \text{ at }  (w_2,\bar w_2)=(x-t_p-i\epsilon,x+t_p+i\epsilon),
 \eeq
  where $w=x+i t_E$, $\bar w=x-it_E$. When mapped to the plane with the map $z=e^{\frac{2\pi}{\beta}w}$, we can see that the two points $P^\infty_R$ and $P^\infty_L$ are secretly the same and we get a single cut running from 
 \beq
 (z_a=e^{-\frac{2\pi}{\beta}t},\bar z_a=e^{\frac{2\pi}{\beta}t}) \;\text{ to }\; (z_b=-e^{\frac{2\pi}{\beta}t},\bar z_b=-e^{-\frac{2\pi}{\beta}t}),
 \eeq
 and operator insertions at
\bea
\label{eq:Vinsertions}
V: && z_1&=z_* e^{-i\frac{2\pi}{\beta}\epsilon}, && \bar z_1&= \bar z_* e^{i\frac{2\pi}{\beta}\epsilon},\\
V^\dagger: && z_2&=z_* e^{i\frac{2\pi}{\beta}\epsilon}, && \bar z_2&= \bar z_* e^{-i\frac{2\pi}{\beta}\epsilon},
\eea
where $z_*=e^{\frac{2\pi}{\beta}(x-t_p)}$, $\bar z_*=e^{\frac{2\pi}{\beta}(x+t_p)}$. 

We will use the replica trick of \cite{Lashkari:2015dia} for the relative entropy
\beq
S(\rho_V||\rho_W) = -\lim_{n\rightarrow 1} \partial_n \log \frac{\text{Tr}\rho_V^n}{\text{Tr}\rho_V \rho_W^{n-1}}.
\eeq 
We now want to compute $\text{Tr}\rho_V^n$ and $\text{Tr}\rho_V \rho_W^{n-1}$, which are given by the Euclidean path integral on an $n$-sheeted cylinder with appropriate operator insertions on each sheet. We do this by uniformizing to the plane with the map $\tilde z=\left(\frac{z-e^{-\frac{2\pi}{\beta}t}}{z+e^{\frac{2\pi}{\beta}t}}\right)^{1/n}$ and $\bar {\tilde z}=\left(\frac{\bar z-e^{\frac{2\pi}{\beta}t}}{\bar z+e^{-\frac{2\pi}{\beta}t}}\right)^{1/n}$. We get the following insertion positions
\beq
\label{eq:planeoperatorpositions}
 z_{1,2;k} = e^{\frac{2\pi i k}{n}} e^{-\frac{2\pi}{\beta n} t}\left( \frac{\sinh{\frac{\pi (w_{1,2}+t)}{\beta}}}{\cosh{\frac{\pi( w_{1,2}-t)}{\beta}} } \right)^{\frac{1}{n}}, \;\;\;\;\; \bar { z}_{1,2;k}= e^{-\frac{2\pi i k}{n}} e^{\frac{2\pi}{\beta n} t}\left( \frac{\sinh{\frac{\pi (\bar w_{1,2}-t)}{\beta}}}{\cosh{\frac{\pi( \bar w_{1,2}+t)}{\beta}} } \right)^{\frac{1}{n}}.
\eeq
We have dropped the tilde from these insertion points, to ease the notation. It should be understood that a coordinate with a subscript $k$ means that the coordinate is on the uniformized plane. The quantities of interest are then computed by the following correlation functions on the plane
\bea
\label{eq:replicacorr}
\frac{\text{Tr}\rho_V^n}{\text{Tr}\rho_{TFD}^n} & = \frac{\langle \prod_{k=1}^n  V( z_{1;k},\bar{ z}_{1;k})V^\dagger( z_{2;k},\bar{ z}_{2;k}) \rangle}{\prod_{k=1}^n \langle V( z_{1;k},\bar{ z}_{1;k})V^\dagger( z_{2;k},\bar{ z}_{2;k}) \rangle}, \\
\frac{\text{Tr}\rho_V \rho_W^{n-1}}{\text{Tr}\rho_{TFD}^n} & = \frac{\langle V( z_{1;1},\bar{ z}_{1;1})V^\dagger( z_{2;1},\bar{ z}_{2;1}) \prod_{k=2}^n  W( z_{1;k},\bar{ z}_{1;k})W^\dagger( z_{2;k},\bar{ z}_{2;k}) \rangle}{\langle V( z_{1;1},\bar{ z}_{1;1})V^\dagger( z_{2;1},\bar{ z}_{2;1}) \rangle\prod_{k=2}^n \langle W( z_{1;k},\bar{ z}_{1;k})W^\dagger( z_{2;k},\bar{ z}_{2;k}) \rangle}.
\eea
Here, $\rho_{TFD}$ is the density matrix for the TFD state, without operator insertions.

\subsubsection{Small $\epsilon$ limits}
\label{eq:OPElimits}
\begin{figure}[h!]
\centering
\includegraphics[width=0.5\textwidth]{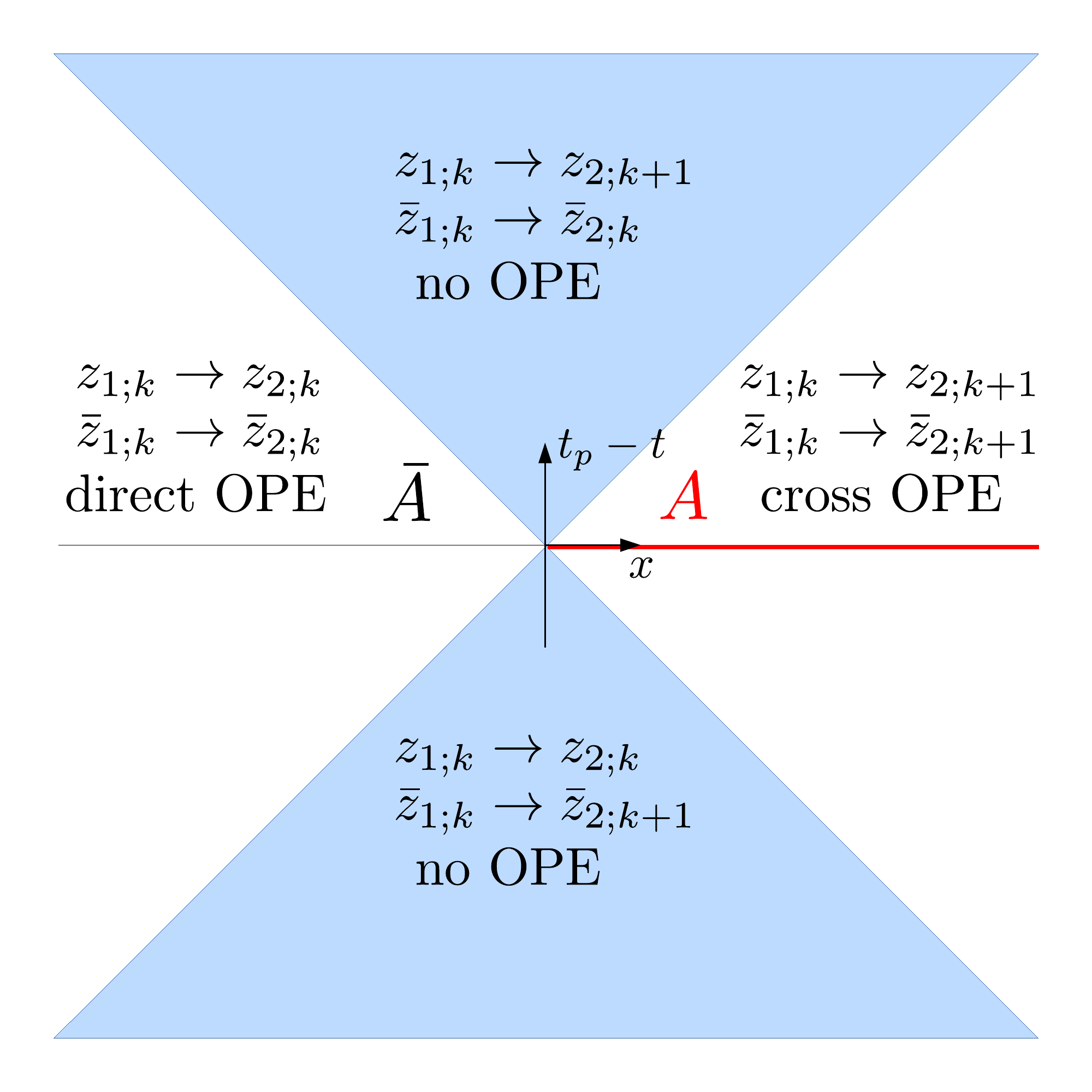}
\caption{Regions of the operator insertion $(x,t_p)$ relative to the endpoint of the subsystem $(0,t)$ on the right CFT, and the respective $\epsilon \rightarrow 0$ limit of the positions of the operator insertions on the replica manifold.}
\label{fig:1}
\end{figure}

The insertion points \eqref{eq:Vinsertions} approach each other in a particular way when we take $\epsilon \rightarrow 0$. The physically distinct cases are controlled by the signs of
\beq
\zeta_{1,2} =  \text{Re} \left(\frac{\sinh{\frac{\pi (w_{1,2}+t)}{\beta}}}{\cosh{\frac{\pi( w_{1,2}-t)}{\beta}} } \right), \;\;\;\;\; \bar {\zeta}_{1,2}= \text{Re}\left( \frac{\sinh{\frac{\pi (\bar w_{1,2}-t)}{\beta}}}{\cosh{\frac{\pi( \bar w_{1,2}+t)}{\beta}} } \right),
\eeq
because when the argument of the $n$th root in \eqref{eq:planeoperatorpositions} is negative, the insertion point picks up an extra factor of $e^{\pm i \frac{\pi}{n}}$, where the sign depends on the sign of the $i\epsilon$ shift in \eqref{eq:Vinsertions}. This difference corresponds to crossing the cut once. The signs of $\zeta_{1,2}, {\bar \zeta}_{1,2}$ are controlled by the causal relationship between the operator insertion point and the endpoint of the subsystem, see Fig. \ref{fig:1}. There are the following cases.

\begin{itemize}
\item
When $|x|>|t-t_p|$, we have either both $\zeta_{1,2}>0$ and ${\bar \zeta}_{1,2}>0$ (when $x>t_p-t>-x$), so the argument of the $n$th root is positive, or both $\zeta_{1,2}<0$ and ${\bar \zeta}_{1,2}<0$ (when $x<t_p-t<-x$), so the argument of the root is negative.  In the $x>t_p-t>-x$ case we have an OPE limit as $\epsilon \rightarrow 0$
\bea
z_{1;k} \rightarrow z_{2;k} && \bar z_{1;k} \rightarrow \bar z_{2;k}.
\eea
This situation is analogous to a small subsystem limit in the setup of globally excited states considered in \cite{Sarosi:2016oks,Sarosi:2016atx}, with roughly $\epsilon$ playing the role of the size of the subsystem. It follows that the relative entropy vanishes as $\epsilon \rightarrow 0$. Physically, this is because the local operator insertion is in the causal domain of dependence of the \textit{traced out} region. Therefore, the RDMs of states with such insertions are the same and we expect the relative entropy to indeed vanish as $\epsilon \rightarrow 0$. This situation is tractable entirely with the OPE, we give a summary of how the relative entropy behaves in Appendix \ref{app:outofcausal}.

 In the case $x<t_p-t<-x$, half of the operators actually cross the cut once more and we have
\bea
z_{1;k} \rightarrow z_{2;k+1} && \bar z_{1;k} \rightarrow \bar z_{2;k+1}.
\eea
This situation is analogous to a large subsystem size limit for globally excited states, therefore the relative entropy must diverge as $\epsilon \rightarrow 0$ (this is because we are comparing pure states in this limit). Physically, this is when the insertion point is in the causal domain of dependence of the region that we are not tracing over, so it is natural that the relative entropy does not vanish as $\epsilon \rightarrow 0$. We mention here, that for the correlators \eqref{eq:replicacorr}, which compute the replica relative entropy, this limit is still an OPE limit and we can calculate their expansion in $\epsilon$. However, we cannot take the analytic continuation because of a singularity in the complex $\epsilon$ plane that moves to the point that we are expanding around ($\epsilon=0$ in the present context) when $n\rightarrow 1$. This is a feature of correlation functions that break replica symmetry. It is presently unclear to us if there is a way around this obstacle.
\item
In the other case, when $|x|<|t-t_p|$, i.e. the insertion is in causal contact with the endpoint of the subsystem, we have either
\bea
\zeta_{1,2}<0, && {\bar \zeta}_{1,2}>0, \;\;\; \text{when}\;\; t_p-t>|x|,
\eea
 or
\bea
\zeta_{1,2}>0, && {\bar \zeta}_{1,2}<0, \;\;\; \text{when} \;\;t_p-t<-|x|.
\eea
This is a weird situation, as the different  chiralities of the operator insertions appear to be on different sheets as $\epsilon \rightarrow 0$. In the first case we have 
\bea
z_{1,k} &\rightarrow z_{2,k+1}\\
\bar z_{1,k} &\rightarrow \bar z_{2,k},
\eea
while in the second case we have
\bea
z_{1,k} &\rightarrow z_{2,k}\\
\bar z_{1,k} &\rightarrow \bar z_{2,k+1}.
\eea
This is not an OPE limit, instead it is a very similar limit as the one considered in \cite{Asplund:2015eha} in the context of entanglement scrambling. In the case of the replica symmetry preserving correlation function, we will soon see that this corresponds to a Regge limit in the cyclic orbifold theory.
\end{itemize}
 We summarize the above cases on Fig. \ref{fig:1}.
 
\subsubsection{Chaos and the late time limit}
\label{sec:chaoslatetime}
Here we argue that in the case when $|x|<|t-t_p|$, i.e. the operator insertion is in the future or past lightcone of the endpoint of the subsystem, the replica relative entropy is sensitive to the integrability of the CFT. 
Consider for example the case $t_p-t<-|x|$, when we have $z_{1,k} \rightarrow z_{2,k}$ and $\bar z_{1,k} \rightarrow \bar z_{2,k+1}$. The point is that the way these coordinates approach each other has a hierarchy. For $x=0$ one has
\beq
\left| \frac{z_{1,k} - z_{2,k}}{\bar z_{1,k} - \bar z_{2,k+1}} \right| = e^{-\frac{4\pi}{\beta n} t}.
\eeq
We see that for $t$ large enough, this is small and we are effectively first taking $z_{1,k} \rightarrow z_{2,k}$ and then $\bar z_{1,k} \rightarrow \bar z_{2,k+1}$.\footnote{For $t_p-t>|x|$, we can get such a hierarchy for $t$ largely negative, in that case we first take $\bar z_{1,k} \rightarrow \bar z_{2,k}$ and then $ z_{1,k} \rightarrow  z_{2,k+1}$.} The effect of first taking $z_{1,k} \rightarrow z_{2,k}$ is to project to chiral operators $h=0$ in the $V \times V^\dagger$ OPE channels of the correlators \eqref{eq:replicacorr}. This means that in this limit the entire correlator is fixed by the chiral algebra of the CFT, so it is linked to the amount of symmetries that the CFT has. This is very similar to the way the depth of the quasiparticle dip works in the two interval entanglement entropy of a quench state \cite{Asplund:2015eha}.

\subsection{The modular Hamiltonian part}
\label{sec:modHam}
After the general discussion of the previous section, we are now going to restrict to the case when one of the states is the unperturbed thermofield double.\footnote{We will further discuss the relative entropy between two perturbed states in appendix \ref{sec:distinctstates}.} In this case, we can calculate the relative entropy as
\beq
\label{eq:relentmodham}
S(\rho_W || \rho_{TFD}) =  \langle W|K_{TFD}|W \rangle -\big[S(\rho_W)-S(\rho_{TFD})\big], \quad S(\rho)\equiv -\text{Tr}\rho \log \rho,
\eeq
where $K_{TFD}=-\log \rho_{TFD}+\alpha$ is the modular Hamiltonian for the thermofield double state where we have fixed the number $\alpha$ so that $ \langle TFD|K_{TFD}|TFD \rangle=0$. In the present subsection we evaluate $\langle W|K_{TFD}|W \rangle$, which is entirely fixed by kinematics.

\begin{figure}[h!]
\centering
\includegraphics[width=0.4\textwidth]{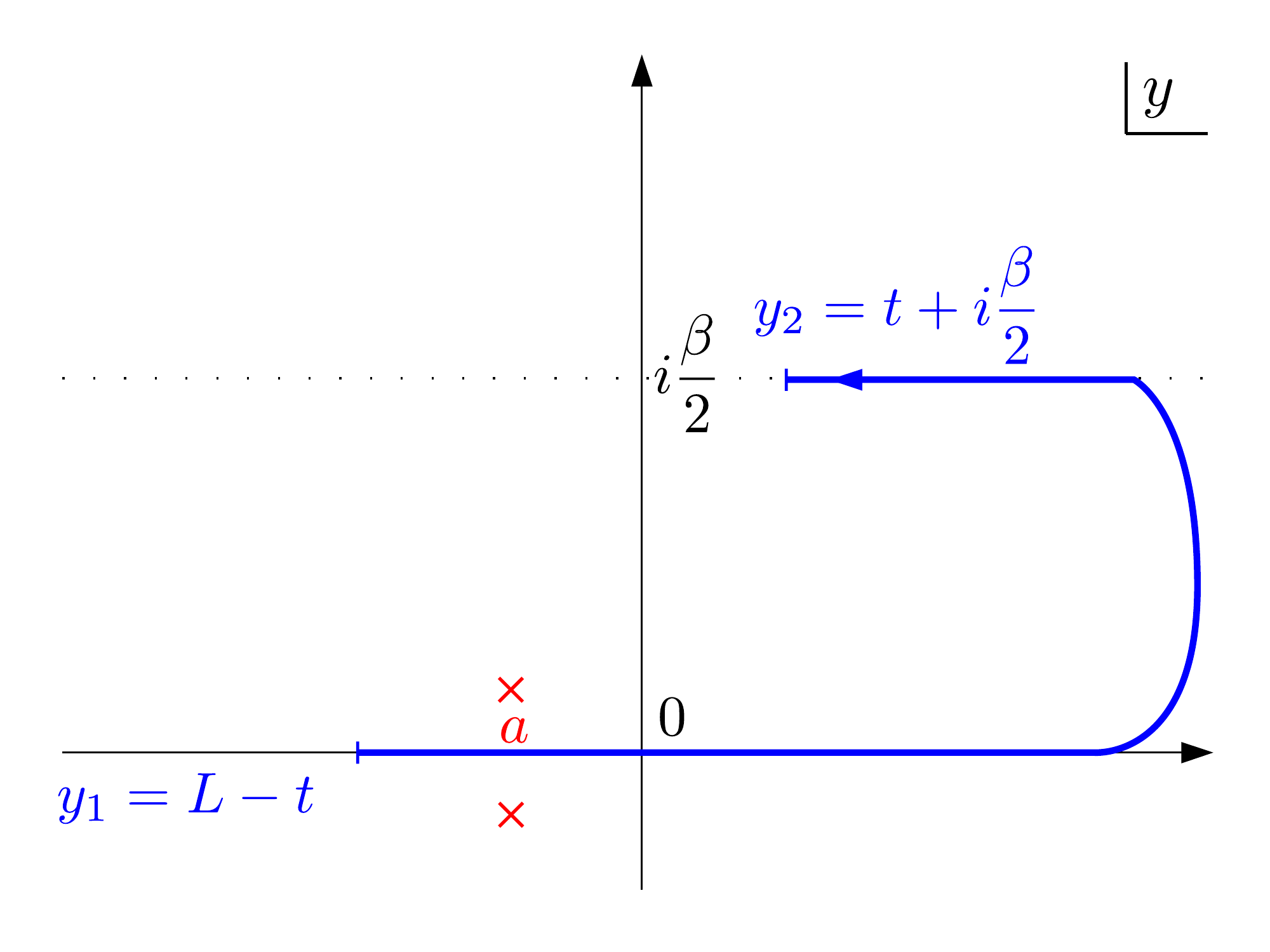} \includegraphics[width=0.4\textwidth]{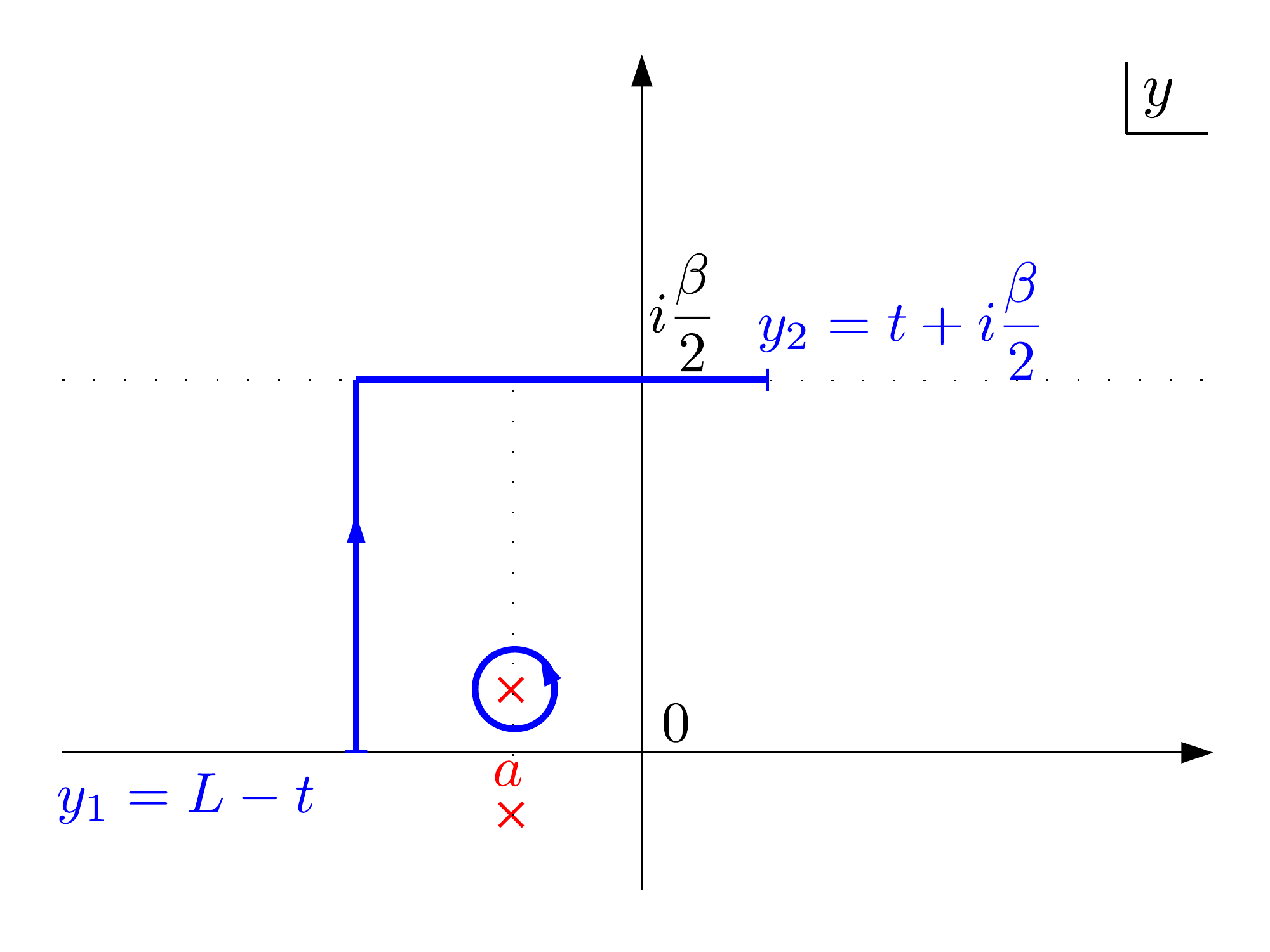}
\caption{Left: the modular Hamiltonian contour in \eqref{eq:modhamcontour}. Right: The deformed contour when $a>L-t$. When $a<L-t$, the contour stays clear of the poles at $a \pm i\epsilon$ and we can deform it into a finite length one without picking up any residue.}
\label{fig:3}
\end{figure}

In this section, we will consider the slightly generalized subsystem consisting of the union of the intervals
\beq
A:(t,x>L) \;\;\;\text{ and }\;\;\; B:(-t-i\beta/2,x>0),
\eeq
i.e. a half line with adjustable end point $L$ on right CFT and a half line on left CFT. This is still effectively a single interval setup and in terms of light cone coordinates the endpoints are 
\beq
(y_1,\bar y_1)=(L-t,L+t)\;\;\;\text{ and }\;\;\; (y_2,\bar y_2)=(t+i\beta/2,-t-i\beta/2).
\eeq
 The left moving part of the modular Hamiltonian can be obtained by the conformal map $z=e^{\frac{2\pi}{\beta}y}$ from the vacuum modular Hamiltonian on the plane
\be
K_{z_{1},z_{2}} =\int^{z_{2}}_{z_{1}} \f{(z_{2}-z)(z-z_{1})}{z_{2}-z_{1}} T(z) dz
\ee
 along the lines of \cite{Cardy:2016fqc}. It is given by 
\beq
\label{eq:modhamintegral0}
K_L =\frac{\beta}{\pi}\int_C dy \frac{\cosh \pi \frac{y-t}{\beta} \sinh \pi \frac{t+y-L}{\beta}}{\cosh \pi \frac{L-2 t}{\beta}}T(y),
\eeq
where the contour $C$ runs from $y_1=L-t$ along the line Im$y=0$ to $y=\infty$, where it turns up to $y=\infty+i\beta/2$ and runs back to $y_2=i\beta/2$, see the left of Fig. \ref{fig:3}. This is just an integral along the subsystem shown on Fig. \ref{fig:setup}. The right moving part is obtained by complex conjugation of this contour, and a replacement $t\rightarrow -t$ in both the endpoint positions and the integrand. The complete modular Hamiltonian is 
\beq
K=-K_L-K_R,
\eeq
because $T_{00}=-\frac{1}{2\pi}(T+\bar T)$.
The stress tensor expectation value $ \la \Psi | T(y)|  \Psi \ra$ in the state of interest \eqref{eq:Vinsertions} is computed from the three point function, 
\be 
\f{\la W(w_{1}) T (y) W(w_{2}) \ra}{\la W(w_{1})W(w_{2}) \ra } =\f{h_{W} \left( \frac{\beta}{\pi}\sinh \pi\f{w_1-w_2}{\beta}\right)^2}{ \left( \frac{\beta}{\pi}\sinh \pi\f{w_1-y}{\beta}\right)^2 \left( \frac{\beta}{\pi}\sinh \pi \f{y-w_{2}}{\beta}\right)^2},
\ee
here we employed the normalization of $W$, $\la W(\infty) W(0) \ra$=1. 
In our set up $w_{1},w_{2}$ are given by 
\be 
w_{1} = x-t_p+i\epsilon, \quad w_{2} =x-t_p-i \epsilon, 
\ee
With the aid of this, we can write the explicit expression for the expectation value $\langle K_L \rangle \equiv \langle W|K_L|W\rangle$ of \eqref{eq:modhamintegral0}
\beq
\label{eq:modhamcontour}
\langle K_L \rangle = \frac{4\pi h_W}{\beta}\sin^2 \frac{2\pi \epsilon}{\beta}\int_C  dy\frac{\cosh \pi \frac{y-t}{\beta} \sinh \pi \frac{t+y-L}{\beta}}{\cosh \pi \frac{L-2 t}{\beta}} \frac{-1}{\left( \cos \frac{2\pi \epsilon}{\beta}-\cosh \frac{2\pi}{\beta}(y-a)\right)^2},
\eeq
with $a=x-t_p$ for the present left moving case, while the right moving case is obtained again by complex conjugation and $t\rightarrow -t, t_p \rightarrow -t_p$.
The integrand has poles at $y=a \pm i\epsilon + i\beta k$, $k\in\mathbb{Z}$. We have two possible cases: 
\begin{itemize}
\item In case $L-t>a=x-t_p$ the contour stays clear of the vicinity of any poles. We may send $\cos \frac{2\pi \epsilon}{\beta} \rightarrow 1$ in the integrand and we can deform the contour into the sum of two finite straight pieces, each of which on the integrand stays finite.
The result in this case is clearly finite, therefore we have $\langle K \rangle \sim \epsilon^2$ coming from the prefactor of the integral. 
\item When $L-t<a=x-t_p$, we can deform the contour through the pole at $y=a+i\epsilon$ to obtain again a finite length contour which stays clear of singularities as $\epsilon \rightarrow 0$, see the right of Fig. \ref{fig:3}. The price we pay for this is that we pick up a residue at  $y=a+i\epsilon$. Therefore, in this case we have
\bea
\langle K_L \rangle &=\frac{4\pi h_W}{\beta} \Big\lbrace \left[\frac{\beta}{2} \frac{\sinh \frac{\pi(L-2t)}{\beta}-\sinh \frac{\pi(2a-L)}{\beta}}{\sin \frac{2\pi \epsilon}{\beta}\cosh\frac{\pi(L-2t)}{\beta}} + O(\epsilon)\right] \\ &+\sin^2 \frac{2\pi \epsilon}{\beta}\Big(\text{finite}\Big) \Big\rbrace.
\eea
Notice that the condition $L-t<a$ guarantees this to be negative. 
Note that the left moving part of the total energy of the state can be obtained by taking $t\rightarrow \infty$ in this formula, and it is $E_W \sim h_W/\sin \frac{2\pi \epsilon}{\beta}$.
To obtain the right moving part of the modular Hamiltonian expectation value $\langle K_R \rangle$, we set $t\rightarrow -t, t_p \rightarrow -t_p$ (the results are obviously real), and we need $t+L<\bar a = x+t_p$ to obtain a nonvanishing result as $\epsilon \rightarrow 0$. 
\end{itemize}

We will now restrict again to the symmetric interval case $L=0$.
We have seen that the left moving part contributes only when $-t<x-t_p$ while the right moving part contributes only when $t<x+t_p$. These two cases correspond to the union of the right and bottom wedges for the case $-t<x-t_p$ and the union of the right and top wedges for $t<x+t_p$ on Fig. \ref{fig:1}. There are the following cases.

\begin{itemize}
\item {\bf Top} The perturbing operator $W$ is inserted in the causal future of the endpoint of the subsystem. We see that only the right moving part contributes to nonvanishing pieces in $\epsilon$. The total modular Hamiltonian contribution is
\bea
\label{eq:topwedgemodham}
\langle K \rangle = -\langle K_R \rangle = 2h_W \pi  \frac{\sinh \frac{2\pi (x+t_p)}{\beta}-\sinh \frac{2\pi t}{\beta}}{\sin \frac{2\pi \epsilon}{\beta}\cosh\frac{2\pi t}{\beta}} .
\eea
We have $t<x+t_p$ in this region ensuring positivity.
\item {\bf Bottom} The operator is inserted in the causal past of the endpoint of the subsystem. In this case, only the left moving part contributed, giving
\beq
\label{eq:bottomwedgemodham}
\langle K \rangle =-\langle K_L \rangle = 2h_W \pi  \frac{\sinh \frac{2\pi t}{\beta}+\sinh \frac{2\pi (x-t_p)}{\beta}}{\sin \frac{2\pi \epsilon}{\beta}\cosh\frac{2\pi t}{\beta}} 
\eeq
Here, $t>t_p-x$ ensures positivity. We will have in mind a situation when the perturbing operator is inserted at some early time $t_p<0$, $|t_p|\gg1$, while we follow the evolution in $t$.
\item {\bf Right} This is the causal diamond of the subsystem. In this case both the left and right moving parts contribute, giving in total
\bea
\label{eq:rightwedge}
\langle K \rangle &=-\langle K_L \rangle-\langle K_R \rangle
 \\
&=4h_W \pi \frac{\cosh \frac{2\pi t_p}{\beta} \sinh \frac{2\pi x}{\beta}}{\sin \frac{2\pi \epsilon}{\beta}\cosh\frac{2\pi t}{\beta}}.
\eea
Here, $x>0$ ensures positivity.
\item {\bf Left} This is the causal diamond of the complementary subsystem. In this case, neither the left nor the right moving part contributes and the result is
\beq
\langle K \rangle \sim  \epsilon^2.
\eeq
We will not try to evaluate the finite integrals in this case, instead we give a separate treatment of the relative entropy in appendix \ref{app:outofcausal}.
\end{itemize}

\subsection{The entanglement entropy part in general}

The entanglement entropy part in \eqref{eq:relentmodham} in our two sided setup was studied before in \cite{Caputa:2015waa}. Here we review this setup in a slightly different way which makes the connection to the behaviour of OTO correlators more transparent.
To calculate the entanglement entropy in \eqref{eq:relentmodham}, we can use a $\mathbb{Z}_n$ symmetric replica trick, which can be implemented in the orbifold theory $CFT^n/\mathbb{Z}_n$, see e.g. \cite{Calabrese:2009qy,Asplund:2014coa}. In this case, we can write
\beq
\frac{\text{Tr}\rho_W^n}{\text{Tr}\rho_{TFD}^n} = \frac{\langle {\tilde \sigma}_n(z_b,\bar z_b) [W^\dagger]^{\otimes n}(z_1,\bar z_1)\sigma_n(z_a,\bar z_a)W^{\otimes n}(z_2,\bar z_2) \rangle}{\langle {\tilde \sigma}_n(z_b,\bar z_b)\sigma_n(z_a,\bar z_a)\rangle \langle [W^\dagger]^{\otimes n}(z_1,\bar z_1)W^{\otimes n}(z_2,\bar z_2) \rangle}.
\eeq
Here $\sigma_n$ and ${\tilde \sigma}_n$ are the elementary $\mathbb{Z}_n$ twist and anti-twist, and the expectation value is in the theory $CFT^n/\mathbb{Z}_n$. The operator ordering in the nominator reflects the Euclidean cylinder time order for the operators: $W$ and $W^\dagger$ has $\mp \epsilon$, $\sigma_n$ has 0 and ${\tilde \sigma}_n$ has $\beta/2$.

We can use a global conformal map to map these operators to $\infty, 1, u,0$. There is no Jacobian factor coming from this, because it is cancelled by the two-point function factors in the denominator. The cross-ratio is
\bea
\label{eq:crossratio}
u &=\frac{(z_1-z_2)(z_b-z_a)}{(z_2-z_a)(z_1-z_b)}\\ &=\frac{i \sin( \frac{2\pi}{\beta} \epsilon )\cosh (\frac{2\pi}{\beta}t)}{\sinh \frac{\pi}{\beta}(t-t_p+x+i\epsilon) \cosh \frac{\pi}{\beta}(t+t_p-x+i\epsilon)},
\eea
while antiholomorhic cross ratio is
\beq
\bar u=\frac{-i \sin( \frac{2\pi}{\beta} \epsilon) \cosh (\frac{2\pi}{\beta}t)}{\sinh \frac{\pi}{\beta}(t_p+x-t-i\epsilon) \cosh \frac{\pi}{\beta}(t+t_p+x+i\epsilon)},
\eeq
The R\'enyi entropy is then given with the cross-ratios as
\beq
\label{eq:orbifoldcrossratio}
\frac{\text{Tr}\rho_V^n}{\text{Tr}\rho_{TFD}^n} = \frac{\langle W^{\otimes n}(\infty)[W^\dagger]^{\otimes n}(1)\sigma_n(u,\bar u) {\tilde \sigma}_n(0) \rangle}{\langle \sigma_n(u,\bar u) {\tilde \sigma}_n(0) \rangle}.
\eeq
Notice that both cross ratios are small for small $\epsilon$ for all times, as long as $t-t_p+x$ and $t-t_p-x$ are neither close to zero, which means that the operator insertion is not lightlike separated from the endpoint of the subsystem. In the case of lightlike separation, when say $\kappa = t-t_p+x=0$, we have $u \rightarrow 2$ as $\epsilon \rightarrow 0$. Similarly, when $\bar \kappa =x-t+t_p=0$, we have $\bar u \rightarrow 2$. As $\kappa$ changes sign, $u$ encircles $u=1$, going from $0$ to $2$ from bellow and then going back to $0$ from above. Similar statement holds for $\bar u$ and $\bar \kappa$. On the $t_p-t=0$ section, we are in Euclidean signature, and $\bar u$ is the complex conjugate of $u$. The correlator is single valued on this section, and it is thus given by the usual Euclidean conformal four point function. We now need to track how the cross ratios move around $1$ as we move away from here, as we expect it to cross a branch cut whenever $\kappa$ or $\bar \kappa$ changes sign. We have the following situations, depending on which wedge the operator insertion is in on Fig. \ref{fig:1}:

\begin{itemize}
\item {\bf Left} Here $\kappa<0,\bar \kappa<0$. We take this to be the reference region, as we have seen in sec. \ref{eq:OPElimits} that this should correspond to a direct OPE limit and the entanglement entropy must vanish as $\epsilon \rightarrow 0$ due to causality reasons. Therefore, in this region we take the limit $u\rightarrow 0$, $\bar u \rightarrow 0$ directly in the Euclidean correlator.
\item {\bf Right} Here $\kappa>0,\bar \kappa>0$ and this corresponds to doing the analytic continuation
\beq
\label{eq:rightcont}
(1-\bar u) \rightarrow e^{2\pi i}(1-\bar u), \;\;\;\; (1- u) \rightarrow e^{-2\pi i}(1-u),
\eeq
to the result in the left region, and then taking $u\rightarrow 0$, $\bar u \rightarrow 0$. Since we can do this continuation while we stay on the Euclidean section ($\bar u$ is the conjugate of $u$), the correlation function must be single valued and this is an OPE limit.\footnote{The fact that the correlation function is single valued on the Euclidean section follows from crossing symmetry.} This is consistent with the fact that the R\'enyi entropies of the complement subsystem must agree with that of the original.
\item {\bf Top} Here $\kappa<0,\bar \kappa>0$ and this corresponds to doing the analytic continuation
\beq
(1- \bar u) \rightarrow e^{-2\pi i}(1- \bar u),
\eeq
to the result in the left region, and then taking $u\rightarrow 0$, $\bar u \rightarrow 0$. in \eqref{eq:orbifoldcrossratio}. This is the standard OTO continuation of the four point function, which diagnoses scrambling and the Lyapunov behaviour \cite{Roberts:2014ifa,Perlmutter:2016pkf}.\footnote{The standard OTO analytic continuation is $(1-u)\rightarrow e^{-2\pi i}(1-u)$, but we can combine this with the continuation $(1-\bar u) \rightarrow e^{-2\pi i}(1-\bar u)$, $(1- u) \rightarrow e^{2\pi i}(1-u)$, which does nothing, to see that they are equivalent.}
\item {\bf Bottom} Here $\kappa>0,\bar \kappa<0$ and this corresponds to doing the analytic continuation
\beq
(1- u) \rightarrow e^{2\pi i}(1- u),
\eeq
to the result in the left region, and then taking $u\rightarrow 0$, $\bar u \rightarrow 0$. Similarly to the previous point, this is also an OTO continuation.

\end{itemize}

Notice that when neither $\kappa$ nor $\bar \kappa$ are close to zero, the cross ratios are related to the modular Hamiltonian expectation values \eqref{eq:topwedgemodham}, \eqref{eq:bottomwedgemodham} in a simple way
\bea
\label{eq:crossratioandmodularham}
\langle K_L \rangle &=-\frac{4\pi i h_W }{u}, && \langle K_R \rangle &=\frac{4\pi i h_W }{\bar u}.
\eea

\subsection{The entanglement entropy part from large $c$ vacuum block}
\label{sec:vacblockee}
We can use the heavy-heavy-light-light large $c$ Virasoro vacuum block to evaluate these R\'enyi entropies, as done in \cite{Caputa:2015waa}. Note, however, that we will focus here on the shockwave limit, so the expressions will be different. As in \cite{Caputa:2015waa}, we treat $W^{\otimes n}$ as the heavy operator and $\sigma_n$ as the light operator.\footnote{We will ultimately have in mind a situation where $h_W$ is $O(1)$ (opposed to what is considered in \cite{Caputa:2015waa}) and the enhancement comes from taking the shockwave limit in the crossratio \eqref{eq:crossratio}, meaning that we take $e^{\frac{2\pi}{\beta}|t_p|} \sim c$. This kinematic limit allows treating $W^{\otimes n}$ as a heavy operator, similarly as done in \cite{Roberts:2014ifa}. The twist field has weight porportional to $c$ but it also vanishes in the $n\rightarrow 1$ limit so keeping only the leading order in $h_n/c$ suffices to get the entanglement entropy accurately, see \cite{Asplund:2014coa}.} The relevant conformal block is \cite{Fitzpatrick:2014vua}
\beq
\label{eq:vacuumblock}
 \mathcal{F}(\bar u) \approx \left( \frac{\bar u}{1-(1-\bar u)^{1-12  h_W/c}}\right)^{2 h_n},
\eeq
where $h_n=\frac{c}{24}(n-1/n)$ is the conformal weight of the twist operator. Note that $W^{\otimes n}$ has weight $n h_W$ but we cancelled this by using that the orbifold central charge is $cn$. The $u\rightarrow 0$ limit is
\beq
\mathcal{F}(\bar u) \approx \left( \frac{1}{1-\frac{12h_W }{c}}\right)^{2 h_n} \approx 1,
\eeq
given that $h_W/c\ll 1$. On the other hand, doing the continuation $(1- \bar u) \rightarrow e^{-2\pi i}(1-\bar u)$, relevant for the top wedge in Fig. \ref{fig:1}, and then taking $\bar u\rightarrow 0$ leads to a decay such as 
\beq
\label{eq:OTOblock}
\mathcal{F}(\bar u) \approx \left( \frac{1}{1-\frac{24 \pi i h_W }{c \bar u}}\right)^{2 h_n} .
\eeq
This decay is also responsible for the decay of OTO correlators \cite{Roberts:2014ifa}. In our case, however, the smallness of the cross-ratio is controlled by a combination of $\epsilon$ and the time. Writing 
\beq
\label{eq:orbifoldvacuumblock}
\frac{\text{Tr}\rho_W^n}{\text{Tr}\rho_{TFD}^n} \approx \mathcal{F}(u)\bar {\mathcal{F}}(\bar u),
\eeq
we can evaluate the entanglement entropy difference required for the relative entropy as
\beq
S(\rho_W)-S(\rho_{TFD}) = -\partial_n [\mathcal{F}(u) {\mathcal{F}}(\bar u)].
\eeq
The result of this approximation, in the different wedges of Fig. \ref{fig:1}, can be neatly summarized using \eqref{eq:crossratioandmodularham} as
\bea
\label{eq:topwedgeEE}
S(\rho_W)-S(\rho_{TFD}) &= \frac{c}{6}\log \left( 1+ \frac{6}{c} \langle K \rangle\right) ,
\eea
where $\langle K \rangle$ is the $O(\epsilon^{-1})$ part of the modular Hamiltonian expectation value in the corresponding wedge, given in \eqref{eq:topwedgemodham},\eqref{eq:bottomwedgemodham}. The exception from this formula is the right wedge of Fig. \ref{fig:1}, i.e. the domain of dependence of the subsystem. Here, the result must agree with that of the left wedge, because of $S(A)=S(\bar A)$ for pure states. Therefore, the entanglement entropy goes to zero as $\epsilon \rightarrow 0$. On the other hand, we would get \eqref{eq:topwedgeEE} in the right wedge also by naively doing the continuation \eqref{eq:rightcont} to the large $c$ vacuum block. We are not allowed to do this for the following reason. While we know that the continuation \eqref{eq:rightcont} leaves the total correlator invariant, it still changes the OPE channel.\footnote{See e.g. \cite{Maloney:2016kee}, the blocks are single valued on the upper half plane. We thank Henry Maxfield for discussion on this point.} In the right wedge, the analytically continued blocks never dominate the correlator, since the new ``direct" channel gives a vacuum block that does not decay as we take $u,\bar u \rightarrow 0$. This is the reason for formula \eqref{eq:topwedgeEE} not being valid in the right wedge.

Finally, based on the discussion of sec. \ref{sec:chaoslatetime} we expect that this answer remains valid for times $\beta \ll t \ll \beta \log c$ for any large $c$ CFT where the chiral algebra consists of the stress tensor alone, regardless of any sparseness condition on the spectrum.

\section{Discussion of the result}
\label{sec:formdisc}
\subsection{Timescales in the large $c$ Virasoro answer}

Let us focus on the bottom wedge in Fig. \ref{fig:1}. In this case, the large $c$ vacuum block approximation results in a relative entropy (combine \eqref{eq:topwedgemodham} and \eqref{eq:topwedgeEE})
\beq
\label{eq:generalform}
S(\rho_W||\rho_{TFD}) = c \Big[q-\frac{1}{6}\log \big( 1+6 q\big) \Big],
\eeq
with
\beq
q= \frac{1}{c} \langle K \rangle = \frac{2h_W \pi}{c}  \frac{\sinh \frac{2\pi t}{\beta}+\sinh \frac{2\pi (x-t_p)}{\beta}}{\sin \frac{2\pi \epsilon}{\beta}\cosh\frac{2\pi t}{\beta}}.
\eeq
We are interested in the shockwave limit, where the perturbation $W$ is inserted at very early times, $t_p<0$. We will set 
\beq
t_p=-t_W, \;\;\; t_W>0
\eeq
 and take the insertion time to scale with the central charge as 
\beq
\label{eq:twscaling}
e^{\frac{2\pi}{\beta}t_W}\sim c.
\eeq
 Notice that in this limit, the parameter $q$ starts out as order one. As $t$ increases, $q$ exponentially decays and eventually becomes $O(c^{-1})$ at the scrambling time $t\sim \log c$. This is when $1/c$ corrections to the vacuum block become important (or equivalently quantum corrections to the Ryu-Takayanagi formula) and we can no longer trust the result. There is actually another characteristic timescale of the above relative entropy. This is set by
\beq
q \approx \frac{1}{6},
\eeq
which is a time $t\sim \beta \log E_W$, where $E_W\sim h_W/\sin\frac{2\pi}{\beta} \epsilon$, i.e. the total energy of the perturbation. At earlier times, $q \gg \frac{1}{6}$ is large and
\beq
S(\rho_W||\rho_{TFD}) \approx h_W \pi \frac{e^{\frac{2\pi}{\beta}({x+t_W)}}}{(\sin \frac{2\pi \epsilon}{\beta}) \cosh \frac{2\pi t}{\beta}}-\frac{c}{6} \log \left(\frac{6h_W \pi}{c} \frac{e^{\frac{2\pi}{\beta}({x+t_W)}}}{(\sin \frac{2\pi\epsilon}{\beta}) } \right) + O(c^0), \label{eq:mainresult}.
\eeq 
This is an exponential decay in $t$ with exponent $2\pi/\beta$. After these times, $q\ll \frac{1}{6}$ is small and we may expand the logarithm. The linear piece gets cancelled with the modular Hamiltonian piece resulting in
\bea
S(\rho_W||\rho_{TFD}) &\approx 3 c q^2 \\
&\approx 3 c \left( \frac{h_W \pi}{c} \frac{e^{\frac{2\pi}{\beta}({x+t_W)}}}{(\sin \frac{2\pi \epsilon}{\beta}) \cosh \frac{2\pi t}{\beta}} \right)^2,
\eea
where we have again assumed $t_W \gg t$. This is an exponential decay again, but the exponent crossed over to $\frac{4\pi}{\beta}$. This decay then continues until the relative entropy reaches $O(1)$ values and our approximations are no longer valid. This happens at the scrambling time, $q^2\sim c^{-1}$.

\subsection{Relative entropy and the chaos bound}

Let us point out something interesting about this latter regime, i.e. when $q\ll \frac{1}{6}$. This expansion in $q$ is the same as the early time Lyapunov expansion of the OTO continued vacuum block \eqref{eq:OTOblock}
\bea
\mathcal{F}(\bar u) &\approx \left( \frac{1}{1-\frac{24 \pi i h_W }{c \bar u}}\right)^{2 h_n} \\
& \approx 1+ \frac{48  \pi i h_W h_n}{c \bar u} + \cdots,
\eea
which is valid as long as $1 \ll {\bar u}^{-1} \sim \epsilon^{-1} \ll c/h_W$. Notice that writing this we treat $h_n$ as $O(1)$ even though it is proportional to $c$.\footnote{An honest $1/c$ expansion would give $\mathcal{F}(\bar u) \approx e^{2\pi i \frac{h_W}{\bar u}\left(n-\frac{1}{n}\right)}+O(1/c)$, which only differs from the Regge-type expansion at order $(n-1)^2$.} This assumption was also made when we used the formula \eqref{eq:vacuumblock} for the vacuum Virasoro block. 
The fact that the $q$ expansion is formally the same as the early time Lyapunov expansion of the OTOC suggests that the cancellation of the modular Hamiltonian part in \eqref{eq:generalform} for $q\ll \frac{1}{6}$ and therefore the speeding up of the decay could be tied to the saturation of the Maldacena-Shenker-Stanford (MSS) chaos bound. 

Let us give another argument that this is indeed the case. This argument is going to rely on assuming that the four point function \eqref{eq:orbifoldcrossratio} satisfies the MSS chaos bound when continued away from $n$ integer, at least when $\text{Re}n\geq 1$, which we cannot justify, therefore what follows is not a proof. Recall that for a generic large $c$ chaotic CFT in the Regge (or Lyapunov) regime, the normalized four point function is expected to behave as
\beq
\label{eq:largecregge}
F=1+ i f_0 \frac{1}{c} \frac{1}{{\bar u}^{J'-1}} + O(c^{-2}),
\eeq
where $J'$ is the ``effective spin" of the ``Regge pole" \cite{Costa:2012cb,Maldacena:2015waa,Perlmutter:2016pkf}. We usually call the combination $\lambda_L=\frac{2\pi}{ \beta} (J'-1)$ the Lyapunov exponent. The reason for this name is that for OTOC the cross ratio would be $\bar u \sim -e^{-\frac{2\pi}{\beta}t}$ so the above translates to an exponential decay
\beq
|F|=1-\tilde \epsilon e^{\frac{2\pi}{\beta}t(J'-1)} + \cdots,
\eeq
(with $\tilde \epsilon \sim 1/c$), which in some cases is related to the exponential divergence of classical trajectories \cite{larkin1969quasiclassical,Polchinski:2015cea,Maldacena:2015waa}.
The statement of the MSS chaos bound is that $J' \leq 2$. More generally, one has
\beq
\label{eq:MSS}
\frac{1}{1-|F|}\left| \frac{dF}{dt} \right| \leq \frac{2\pi}{\beta}.
\eeq

We would like to translate this bound to our setup, where the same analytic continuation is done for the cross ratios, followed by the same $u,\bar u \rightarrow 0$ limit in the replica four point function \eqref{eq:orbifoldcrossratio} as the one taking the correlator to the Regge limit.\footnote{Note that the Regge and the chaos limits are only the same in two dimensions.} The way $u$ and $\bar u$ approaches zero for the OTOC setup is \cite{Roberts:2014ifa}
\beq
u \sim \alpha e^{\frac{2\pi}{\beta}(x_{OTOC}-t_{OTOC})},\; \;\; \bar u \sim \alpha e^{\frac{2\pi}{\beta}(-x_{OTOC}-t_{OTOC})},
\eeq
for $\alpha$ some complex number, while in our setup we have \eqref{eq:crossratio}, which in the regime where \eqref{eq:twscaling} holds and $\beta \ll t \ll \beta \log c$ reads as
\beq
u \sim \tilde \alpha e^{\frac{2\pi}{\beta}(t-t_W-x)},\; \;\; \bar u \sim \tilde \alpha e^{\frac{2\pi}{\beta}(t-t_W+x)},
\eeq
for $\tilde \alpha$ another complex number. We can therefore identify $t=-t_{OTOC}$, $x=-x_{OTOC}$, $\alpha=\tilde \alpha e^{-\frac{2\pi}{\beta}t_W}$ and we see that we can apply the bound \eqref{eq:MSS} to \eqref{eq:orbifoldcrossratio} directly.

The correlator \eqref{eq:orbifoldcrossratio} of course satisfies the general assumptions of the chaos bound for any $n\in \mathbb{Z}^+$ but since the twists have dimension $O(c)$ it is not clear that they have a Regge limit of the form \eqref{eq:largecregge}. Nevertheless, there is another small parameter, namely $n-1$, in which we can expand:
\beq
\frac{\langle W^{\otimes n}(\infty)[W^\dagger]^{\otimes n}(1)\sigma_n(u,\bar u) {\tilde \sigma}_n(0) \rangle}{\langle \sigma_n(u,\bar u) {\tilde \sigma}_n(0)\rangle} \approx 1 -(n-1) f(t) + \cdots.
\eeq
Assuming that the MSS chaos bound remains valid when we continue away from integer $n$ to $\text{Re}n\geq 1$, we have that
\beq
\label{eq:fbound}
 \frac{|\partial_{t} f|}{\text{Re}f}  \leq \frac{2\pi}{\beta}.
\eeq
Note that $f$ is the entanglement entropy difference $S(\rho_W)-S(\rho_{TFD})$ so it should be real. For the chaos bound to hold we also need $|F|\leq 1$ therefore our assumption requires $f\geq 0$. It is interesting to note that the $f$ coming from \eqref{eq:OTOblock} satisfies this bound for any value of $h_W/c$, and saturates it to leading order in $h_W/c$. This suggests that we have the Regge-like pole in $f$ when we expand in $1/c$, but with a coefficient proportional to $c^0$ instead of $1/c$. 

Now in this limit we can write the relative entropy with \eqref{eq:relentmodham} and \eqref{eq:bottomwedgemodham} as
\beq
S(\rho_W||\rho_{TFD})
\approx 2h_W \pi \frac{e^{\frac{2\pi}{\beta}({x+t_W)}}}{(\sin \frac{2\pi \epsilon}{\beta})}e^{-\frac{2\pi}{\beta} t} -f(t) .
\eeq
We have seen that for the gravitational answer the $e^{-\frac{2\pi}{\beta}t}$ piece gets canceled between $f$ and the modular Hamiltonian piece.
We now see that this cancellation is only possible when the bound \eqref{eq:fbound} is saturated. On the other hand, when there is no cancellation, the bound \eqref{eq:fbound} implies in the regime $\beta \ll t \ll t_W$ that
\beq
\label{eq:decaylowerbound}
-\frac{\partial_t S(\rho_W||\rho_{TFD})}{S(\rho_W||\rho_{TFD})} \geq \frac{2\pi}{\beta}.
\eeq 
This is a \textit{lower} bound on the decay rate which might seem surprising. However, one must keep in mind that it relies on our somewhat bold assumption about the analytic continuation of the chaos bound. We note that for the gravitational answer \eqref{eq:generalform} in the regime $\beta \log h_W/\epsilon \ll t \ll t_W$ we actually have
\beq
-\frac{\partial_t S(\rho_W||\rho_{TFD})}{S(\rho_W||\rho_{TFD})} \sim \frac{4\pi}{\beta}.
\eeq

\subsection{Comment on integrable systems}

For integrable systems (or systems with $\lambda_L$=0), we expect that both the replica four point function and the entanglement entropy is $O(\epsilon)$. This is because the Regge limit of the four point function \eqref{eq:orbifoldcrossratio} is expected to contain only positive powers of the cross-ratio $u$. As a consequence, we expect that the relative entropy is dominated by the modular Hamiltonian piece. For example, for chiral vertex operators $V_\alpha=e^{i\alpha X}$ of the free boson, one may explicitly check that\footnote{Parametrizing $w=e^{ix}$ and using a uniformization map to map the branched correlator to the plane we have
\beq
\frac{\langle V_{-\alpha}^{\otimes n}(\infty) \sigma_n(1)\tilde \sigma_n (w) V_{\alpha}^{\otimes n}(0)  \rangle}{\langle  \sigma_n(1)\tilde \sigma_n (w) \rangle}  = \left[ \frac{2}{n} \sin \pi x \right]^{n \alpha^2} \langle \prod_{k=1}^n V_{-\alpha}(w_k) V_\alpha({\hat w}_k) \rangle,
\eeq
where $w_k=e^{2\pi i \frac{k}{n}}$ and ${\hat w}_k=e^{2\pi i \frac{x+k}{n}}$. Using the known vertex operator $2n$ point function \cite{DiFrancesco:1997nk} gives a cancellation between the correlator and the Jacobian factor. For more on relative entropy for the 2d free boson, see \cite{Lashkari:2015dia,Ruggiero:2016khg,Nakagawa:2017fzo}.}
\beq
\frac{\langle V_{-\alpha}^{\otimes n}(\infty) \sigma_n(1)\tilde \sigma_n (w) V_{\alpha}^{\otimes n}(0)  \rangle}{\langle  \sigma_n(1)\tilde \sigma_n (w) \rangle} =1,
\eeq
so that the entanglement entropy difference is just zero
\beq
S(\rho_{V_\alpha})-S(\rho_{TFD})=0,
\eeq
and the relative entropy is entirely given by the modular Hamiltonian part. Therefore, we expect that for integrable systems, the complete exponentially decaying part has exponent $2\pi /\beta$ until the decay stops.

\subsection{The $t_W$ dependence}
\label{sec:tw}
Let us briefly comment on the dependence of the result \eqref{eq:generalform} on the time of insertion of the perturbing operator $W$. Since the relative entropy measures distinguishability of states, its $t_W$ dependence quantifies how far the perturbed state ended up from the thermofield double if a perturbation was made in the past at time $t_p=-t_W<0$. This has a similar flavor to the butterfly effect even though the relative entropy is not obviously related to phase space trajectories in a classical limit. The intuition is that it is a more refined probe of distinguishability than any kind of correlator, such us the commutator-squared correlators that are used to define the Lyapunov exponent. In this light, it is satisfying to see that \eqref{eq:generalform} has a fairly universal exponentially growing dependence on $t_W$, coming from the modular Hamiltonian expectation value \eqref{eq:bottomwedgemodham} that is the same for any 2d CFT. It would be very interesting if this growth could be used to give an information theoretic understanding of the MSS chaos bound.

We close here with something more modest, by showing that the relative entropy setup that we are considering indeed bounds the magnitude of OTO correlators (but not their time derivatives). This is just a simple application of the quantum version of Pinsker's bound \cite{ohya2004quantum}
\beq
\label{eq:ineq1}
S(\rho_W||\rho_{TFD})\geq \frac{1}{2} ||\rho_W-\rho_{TFD}||^2_1,
\eeq
and the duality identity of $||.||_p$ norms 
\beq
\label{eq:ineq2}
||X||_1 = \sup_Y \left( \text{Tr}(XY^\dagger)/||Y||_\infty \right).
\eeq
We pick $X=\rho_W-\rho_{TFD}$ which lives on the two half lines $L_1 \cup R_1$ on the two sides, and restrict the supremum to operators which factorize, i.e. $Y=U_{L_1}Z_{R_1}$. We imagine $U(0)$ and $Z(0)$ to be bounded operators defined on the $t=0$ slice of a single copy of the CFT, such that $U_{L_1}=U(0)$ has support on the left half line $L_1$ and $Z_{R_1}=Z(i\beta/2)$ has support on the right half line $R_1$. Now
\beq
\text{Tr}(\rho_{TFD} U_{L_1}Z_{R_1})
\eeq
is calculated by a path integral on the cylinder, with an insertion of $U$ on one side and an insertion of $Z$ on the other side. Therefore
\beq
\text{Tr}(\rho_{TFD} U_{L_1}Z_{R_1})=\langle U(0)Z(i\beta/2) \rangle_\beta.
\eeq
Here, $\langle . \rangle_\beta$ denotes a thermal expectation value in a single copy of the CFT. On the other hand,
\beq
\text{Tr}(\rho_W U_{L_1}Z_{R_1})
\eeq
is calculated by the same path integral with extra insertions of the local operator $W$ at time $t_W+i\tau$ and $t_W-i\tau$, therefore
\beq
 \text{Tr}(\rho_W U_{L_1}Z_{R_1}) \sim \langle U(0)Z(i\beta/2)W(t_W+i\tau)W(t_W-i\tau) \rangle_\beta \equiv F(t_W) 
\eeq
The path integral of course gives an Euclidean time ordered correlator, and since $0<\tau<\beta/2$ we see that the operator ordering is $WUWZ$ in this correlator, which is precisely the OTOC that we were after. Taking into account that $\rho_W$ is normalized we actually have
\beq
 \text{Tr}(\rho_W U_{L_1}Z_{R_1}) = \frac{F(t_W) }{\langle W(2i\tau )W(0)\rangle_\beta}
\eeq
We now combine \eqref{eq:ineq1} and \eqref{eq:ineq2} with this to get
\beq
\label{eq:uselessbound}
S(\rho_W||\rho_{TFD})\geq \frac{1}{2} \sup_{U,Z} \lbrace \frac{\langle U(0)Z(i\beta/2) \rangle_\beta^2}{||U||_\infty^2 ||Z||_\infty^2}  \left( 1-\frac{F(t_W)}{F_d}\right)^2\rbrace,
\eeq
where
\beq
F_d = \langle U(0)Z(i\beta/2) \rangle_\beta \langle W(2i\tau )W(0)\rangle_\beta.
\eeq
If $U=Z$ were local operators \cite{Maldacena:2015waa}, then for $t_d \ll t_W \ll t_s$ ($t_d$: collosion time, $t_s$: scrambling time) we would be in the Lyapunov regime 
\beq
F(t)\sim F_d - \epsilon e^{\lambda_L t_W} + \cdots,
\eeq
so the r.h.s of \eqref{eq:uselessbound} would be proportional to $e^{2\lambda_L t_W}$. Of course the above bound is only nontrivial when $U$ and $Z$ are bounded operators, so they cannot be local. We think that this is not a major obstacle as long as $W$ is allowed to be local, because of a semiclassical reasoning: divergence of phase space trajectories is equally well measured by the Poisson bracket $\lbrace e^{-q(t)^2},p(0) \rbrace$ as $\lbrace q(t),p(0) \rbrace$. Still, this bound is clearly not related to the chaos bound as it does not say anything about the rate of change.

\section{Holographic calculation I: Global shocks} 
\label{sec:hologlobal}

In the following two sections, we use holography to calculate the relative entropy in similar setups as in the previous section.  Since the bulk geodesics in 3d gravity are directly related to Virasoro 4 point conformal blocks in 2d CFTs in the large central charge limit \cite{Asplund:2014coa}, we will be doing similar calculations as in the previous section. Nevertheless, these computations provide some generalizations compared to the setup of the previous section and we hope that they give additional bulk insights to the physics of scrambling.
\footnote{We also consider a slightly generalized set up in appendix \ref{app:disjoint}, namely when the subsystem is the disjoint union of two finite intervals, one in the right CFT and the other is in the left. This introduces additional finite size effects.}

The strategy to compute the relative entropy is the following. First we split the relative entropy to the modular Hamiltonian  part and the entanglement entropy part. 

\begin{align}
S(\rho|| \rho_{TFD}) &= {\rm tr} \rho \log \rho- {\rm tr} \rho \log \rho_{TFD}, \nonumber \\
&=\left[{\rm tr}\rho K_{TFD}- {\rm tr} \rho_{TFD} K_{TFD}\right] -\left[ S(\rho) - S(\rho_{TFD}) \right],
\end{align}
where $S(\rho) =-{\rm tr} \rho \log \rho$. We use the Ryu-Takayanagi formula to compute the entropy part, and combine it with the universal modular Hamiltonian result of section \ref{sec:modHam}.   

To begin with, we first calculate the relative entropy between the TFD state and its global perturbation. We consider the translationally invariant state,

\be
| \Psi_{S} \ra= U_{L}(t_{W}) |TFD \ra, \label{eq:globalp}
\ee 
where $ U_{L}(t_{W})$ is a unitary operator on left Hilbert space $\mathcal{H}_{L}$, and $t_{L}=t_{W}$ denotes the time when the operator is inserted.  This state, when the left Hilbert space is traced out,  gives a thermal density matrix on the right Hilbert space. Therefore it can be regarded as one of the black hole microstates for the observers of $\mathcal{H}_{R}$. Note that here we insert the operator on the left CFT, whereas in the previous section we added it on the right. This will of course not modify the result in a relevant way. Killing time runs backwards on the left CFT therefore an early perturbation now corresponds to $t_L=t_W>0$ large and positive.

The reduced density matrix of $|\Psi_{S} \ra$ to the union of the positive half lines on both sides will be denotes with $\rho_S$.

\subsection{Dual geometries} 
Here we briefly review the dual geometries of the two states $| TFD \ra$ and $| \Psi_{S}\ra$ in the CFT, in order to fix the notations.

\subsubsection*{BTZ black hole}

In the Schwarzchild coordinates, the metric of the BTZ black hole is given by 
\be
ds^2= -\f{r^2-R^2}{l^2} dt_{R}^2 + \f{dr^2}{r^2-R^2} +r^2 dy^2, \quad R^2 = 8 G_{N} M l^2.
\ee 
$l$ denotes the AdS radius, $R$ the location of the horizon and the inverse temperature of the black hole is given by 
\be
\beta=\f{2\pi l^2}{R}. 
\ee
It is also convenient to introduce the Kruskal coordinates, in which the black hole metric is
\be
ds^2 =\f{-4l^2 du dv +{R^2}(1-uv)^2 d y ^2}{(1+uv)^2}.
\ee
In this coordinates the AdS boundary is $uv=-1$.  Left (Right) CFT is defined on $ u>0$ ($u<0$) part of the curve.

\noindent
The coordinate transformations between these two  is explained in \cite{Shenker:2013pqa}.  Here we only quote the relevant ones

\begin{eqnarray} 
\f{v-u}{1+uv} &=& \f{\s{r^2-R^2}}{R}\cosh \f{2\pi t_{R}}{\beta} \\
\f{u+v}{1+uv}&=& \f{\s{r^2-R^2}}{R}  \sinh \f{2\pi t_{R}}{\beta}
\end{eqnarray}
The original Schwarzchild coordinates $(t_{R},r)$ only cover the right wedge  $u<0, v>0$  of the geometry.  We  have  analogous coordinates  $(t_{L},r)$ on the left wedge  $u>0, v<0$ just by the shift $t_{L}= t_{R}+i\f{\beta}{2}$. 
The black hole in the Kruskal coordinates is dual to the TFD state (\ref{eq:TFDs}).

\hspace{0.3cm}

\subsubsection*{Black hole with a shock wave}

Imagine that we send a null shock wave from the left boundary at $t_{L} =t_{W}$ with energy $E$.    The trajectory of the shock is  $u=e^{-\f{2\pi}{\beta} t_{W}} $.  The shock wave is highly blue shifted near the black hole horizon, and its backreaction  is non-negligible at the horizon. In the limit, 
\be 
E \rightarrow 0, \; t_{W} \rightarrow \infty \quad  {\rm with}  \quad \alpha \equiv  \f{E}{4M}e^{\f{2\pi}{\beta} t_{W}}  \quad {\rm kept\; fixed},  \label{eq:midya}
\ee
the backreacted metric is \cite{Shenker:2013pqa} 
\be
ds^2 = \f{ -4l^2 du dv +R^2 \big[1-u(v+ \alpha \theta(u))\big]^2 dy^2 }{\big[1+u(v+ \alpha \theta(u))\big]^2},
\ee
where $\theta(u)$ denotes the step function.  This geometry is constructed by gluing two  BTZ black holes with mass $M+E$ and $M$ at the location of the shock wave, and taking the limit (\ref{eq:midya}).  This geometry is dual to (\ref{eq:globalp}), a TFD state perturbed by a unitary transformation $ U_{L}(t_{W})$ at  $t_{L} = t_{W}$  in the left CFT.

\subsection{Holographic calculations of the entanglement entropy}

The holographic entropy is computed by the Ryu-Takayanagi formula 

\be 
S= \f{A(\gamma_{A})}{4G_{N}}, 
\ee
where $A$ is the length of the bulk geodesic $\gamma_{A}$ connecting two end points of the subsystem $A$.  We 
take the same subsystem as in the previous section.  
In the BTZ black hole, the result is given by 
\be
S(\rho_{TFD})  = \f{c}{3}\log \f{r_{\infty}}{R} + \f{c}{3}\log \left[ \cosh \f{2\pi t}{\beta}\right]
\ee
where $r_{\infty}$ denotes the UV cut off.  
In the BTZ black hole with a global shockwave we can also calculate the length of the geodesic, just by  gluing two geodesics in each BTZ black hole ending on the null surface $u=0$, and minimizing the length with respect to the location of the end point.  The result is given by \cite{Shenker:2013pqa}
\be
S(\rho_{S}) = \f{c}{3}\log \f{r_{\infty}}{R} + \f{c}{3}\log \left[ \cosh \f{2\pi t}{\beta} + \f{\alpha}{2} \right]
\ee
$\alpha$ is defined in (\ref{eq:midya}).  

\subsection{Evaluation of modular Hamiltonian part}

Next we compute the modular Hamiltonian expectation value ${\rm tr} \rho_{S} K_{TFD}$.  Notice that
\be 
{\rm tr} \left[ T_{00} (t_{L},x) \rho_{S} \right] = \begin{cases}
\f{E}{2\pi}&  t_{L} \leq t_{W} \\
0& t_{L} \geq t_{W}
 \end{cases}, 
\qquad {\rm tr} \left[T_{00} (t_{R},x) \;\rho_{S} \right]=0,
\ee
since the shockwave is sent in at $t_L=t_W$ and there is never any stress energy on the right. Using $T_{00}=-\frac{1}{2\pi}(T+\bar T)$
 and  the contour prescription described in section \ref{sec:modHam}, adapted so that the perturbation is now made on the left, we have that the modular Hamiltonian expectation value is
\begin{align} 
{\rm tr}  K_{TFD}  \rho_{S}  
 &= \f{\beta E}{2\pi \cosh \f{2\pi t}{\beta} }  \int^{t_{W} +\f{i\beta}{2}}_{-t+\f{i\beta}{2}}\left[\sinh \f{2\pi t}{\beta} -\sinh \f{2\pi y}{\beta} \right] dy  \\[+10pt]
&\rightarrow  \left(\f{\beta}{2\pi} \right)^2 \f{2M \alpha}{ \cosh \f{2\pi t}{\beta}}
\end{align}
in the shockwave limit (\ref{eq:midya}).  By using the relation between the mass and temperature of the black hole
\be 
M= \f{c}{12} \left( \f{2\pi}{\beta}\right)^2
\ee
We have 
\be 
{\rm tr}  K_{(c)} \rho_{S}= \f{c}{6} \f{\alpha}{ \cosh \f{2\pi t}{\beta}}
\ee
This expression is plausible since it obeys the first law  relation 
\be 
\delta S = {\rm tr}  K_{TFD}  \; \rho_{S} + O(\alpha^2)
\ee

\subsection{Relative entropy}

Putting together the pieces, the final result is
\be
S(\rho_{S} || \rho_{TFD}) =\f{c}{6} \f{\alpha}{ \cosh \f{2\pi t}{\beta}}  -  \f{c}{3}\left(\log \left[ \cosh \f{2\pi t}{\beta}+\f{\alpha}{2} \right]- \log  \cosh \f{2\pi t}{\beta}\right).
\ee
Notice that this has the form \eqref{eq:generalform} with the replacement
\beq
c_{\rm there}=2c_{\rm here}, \;\;\;\; q=\f{1}{12} \f{\alpha}{ \cosh \f{2\pi t}{\beta}}, 
\eeq
and hence the time dependence has qualitatively the same behavior as for the local perturbations.

\section{Holographic calculations II: localized shocks} 
\label{sec:hololocal}

In this section we generalize the above holographic calculation to the cases where the perturbations are localized along the spatial direction. This is the same setup as the one one considered in section \ref{section:CFTpart}, where we have obtained the answer with large $c$ vacuum block techniques. Therefore, here we move quickly and only summarize the dual geometry as well as its properties and the resulting relative entropy.

We consider the state realized by an insertion of a local  primary operator $W$, which was defined in (\ref{eq:TFDp})
\be 
| \Psi \ra= W( t_{W} +i \tau, x) | TFD\ra, 
\ee
where $x$ is the location of the insertion, and we have an Euclidean shift   in the timelike direction $ t_{W} \rightarrow t_{W} +i \tau$ to make the state normalizable.  We will set $\tau=\frac{\beta}{2} - \epsilon$, so that the primary is located at the left CFT.  We denote the conformal dimension of the primary by $h_{W}$. 

\subsection{Dual gravity geometry}

The metric of the localized shock is given by\footnote{ In this section we set $\beta=2\pi$} \cite{Roberts:2014ifa,Roberts:2014isa} 
\be 
ds^2 = -\f{4}{(1+uv)^2} dudv + \left(\f{1-uv}{1+uv} \right)^2 dy^{2} +4\delta (u) h(y) du^2
\ee
with 
\be
h(y) =2 \pi G_{N} P e^{-|y-x|}.
\ee
This metric is a solution of Einstein equations in the presence of the bulk stress tensor, 
\be
T_{uu} (u,v,y) = P \delta(u) \delta (y-x).
\ee
By evaluating  $\la \Psi |T_{uu} | \Psi \ra$, one can fix the coefficient 
\be 
P = \f{2h_{W} }{\sin \tau} \;e^{t_{W}}
\ee
This metric can be constructed by gluing two BTZ's at $u=0$ with a shift,
\be 
\delta v  = h(y).
\ee

\subsection{The geodesic length}  

We now calculate the length of the geodesic which is  starting from $ P_{1} :(t,y,r)=( -t +i\f{\beta}{2}, 0, r_{\infty}) $  and ending at $P_{4} :(t,y,r)=( t, 0, r_{\infty}) $. 
This was done in \cite{Roberts:2014ifa}, which we slightly generalize here. The way to calculate this is very similar to the spherically symmetric shock case. We first consider the length of the geodesic $d(P_{1}, P_{2})$ starting from $P_{1}$ and ending at the horizon $ P_{2} (u,v, y) =(0,v+h(y), x)$, as well as $d (P_{3}, P_{4})$ with  $P_{3} = (u,v,y) =(0, v,y)$.

The sum of these geodesic length is given by 
\be
d(v,y)=  2\log \f{r_{\infty}}{R} + \log \left[ \cosh y -v e^{-t} \right] + \log \left[ \cosh y +(v+h(y))e^{t} \right].
\ee 
We then extremalize the function $d(v,y)$ with respect to $v$ and $y$. The resulting holographic entanglement entropy is
\begin{align}
 S_{EE} &=\f{c}{3} \log \cosh t + \f{c}{6}\log \left[ 1+ \f{h(0)}{\cosh t} \right]  \\ 
&=\f{c}{3} \log \cosh t + \f{c}{6}\log \left[ 1+ \f{1}{\cosh t} \left(\f{6\pi h_{W}}{c \sin \tau} \right) e^{t_{W}+x} \right]  \label{eq:sEEshock}
\end{align} 
here $c=3l/2G_{N}$.  
This gives the same entanglement entropy as in \eqref{eq:topwedgeEE}, provided we are in the shockwave limit $e^{t_W}\sim c \gg 1$.
Combining this with the universal results for the modular Hamiltonian expectation value in sec. \ref{sec:modHam}, we obtain the relative entropy $S(\rho_W||\rho_{TFD})$.

\section{Spin chains}
\label{sec:spinchains}
In this section, we present numerical results on the relative entropy in spin chains.
We simulate the localized shock studied in the previous section in a lattice spin chain  
and numerically calculate the relative entropy.
We find an exponential decay of the relative entropy in both integrable and non-integrable spin chains
at early times.
The relative entropy remains small after the decay in non-integrable cases while
weak revivals are observed in integrable cases.
Furthermore, the decay rate seems to be proportional to the temperature, which is consistent with the conformal field theory results.
We also study the $t_W$-dependence of the relative entropy and find an algebraic growth instead of the exponential growth discussed in section \ref{sec:tw}.

\subsection{Setup of spin chain}
We consider a spin chain consisting of $N$ sites, 
\be
H = - \sum_{i=1}^N \left( Z_{i} Z_{i+1} + g X_{i} + h Z_{i} \right), \label{eq:SpinModel}
\ee
where $X_i, Y_i$, and $Z_i$ are the Pauli operators acting on site $i$ and the periodic boundary condition
$ X_{N+1}, Y_{N+1}, Z_{N+1} = X_1, Y_1, Z_1$ is imposed.
When $g \neq 0$, the system is integrable for $h = 0$ and non-integrable for $h \neq 0$.
Throughout this section we choose $g = -1.05, h = 0.5$ for a non-integrable spin chain (the same as in Refs.~\cite{banuls2011strong,Hosur:2015ylk}) and $g=1, h=0$ for an integrable spin chain, i.e., the critical transverse-field Ising chain.

Similarly as Eq.~\eqref{eq:TFDs}, 
the TFD state in spin chains is defined as a pure state of two copies of the spin chain:
\be
 \ket{TFD} := \frac{1}{\sqrt{Z(\beta)}} \sum_{n=1}^{2^N} e^{-\beta E_n /2} \ket{n}_L \ket{n}_R,
\ee
where $L(R)$ denotes the left (right) system, $\ket{n}_{L(R)}$ and $E_n$ are energy eigenstate and energy eigenvalue of the left (right) system,
and $Z(\beta) = \sum_n e^{-\beta E_n}$ is the partition function of the single system at inverse temperature $\beta$.
Time evolution of the TFD state is defined as
\be
 \ket{TFD(t)} = e^{-i (H_L + H_R)t} \ket{TFD}
  =  \frac{1}{\sqrt{Z(\beta)}} \sum_{n} e^{-\beta \frac{E_n}{2} - 2iE_n t} \ket{n}_L \ket{n}_R.
\ee
We note that the time evolution yields (trivial) phase in the wave function.
The (locally) perturbed TFD state is defined as
\be
 \ket{TFD'(t=0, t_W)} := e^{ -i t_W H_L } \left( (Z_x)_\mathrm{left} \otimes \hat{I}_\mathrm{right}\right) e^{ i t_W H_L } \ket{TFD},
\ee
which means that we perturb only the left system by {\it local} Pauli-$Z$ operator
at site $x$ in the past of time $t_W$\footnote{We note that $\ket{TFD'}$ is still normalized because $Z^2=1$.}.  Time evolution of $\ket{TFD'}$ is again
\be
 \ket{TFD'(t, t_W)} = e^{ -i (H_L + H_R) t} \ket{TFD'(t=0, t_W)}.
\ee
We take a subregion $A_1 (A_2)$ as the sites $i=1, \ldots, m$  on the left (right) chain
and define $A= A_1 \cup A_2$.
We consider the $t$- and $t_W$-dependence of the relative entropy between
$\ket{TFD'(t, t_W) }$ and $\ket{TFD(t)}$, namely,
\be
 S(t; t_W) =  S(\rho'_{t_W}(t) \| \rho(t))
\ee
where
\be
 \rho'_{t_W}(t) = \mathrm{Tr}_{\bar{A}} \ket{TFD'(t, t_W)}\bra{TFD'(t, t_W)},  \:\: \rho_0(t) = \mathrm{Tr}_{\bar{A}} \ket{TFD(t)}\bra{TFD(t)}.
\ee
In addition to $S(t; t_W)$, we calculate the mutual information, $I(t; t_W)$,
between the subsystems $A_1$ and $A_2$ of the perturbed TFD state $\ket{TFD'(t, t_W)}$.

\subsection{$t$-dependence of the relative entropy}
First we fix $t_W$ and study $t$-dependence of the relative entropy $S(t; t_W)$.
From the conformal field theory and holographic considerations that we have discussed so far in sections \ref{section:CFTpart},\ref{sec:hologlobal},\ref{sec:hololocal},
we expect this to show an exponential decay in a proper parameter region.
We show that this is indeed the case in the spin chain.
Here we take the size of each subsystem as $m=2$ and set the position of the perturbation as $x=2$.
We vary inverse-temperature $\beta$ and calculate $S(t; t_W)$ by exact diagonalization of the spin chain of $N=8$ sites.

\begin{figure*}
  \includegraphics[width = 8cm]{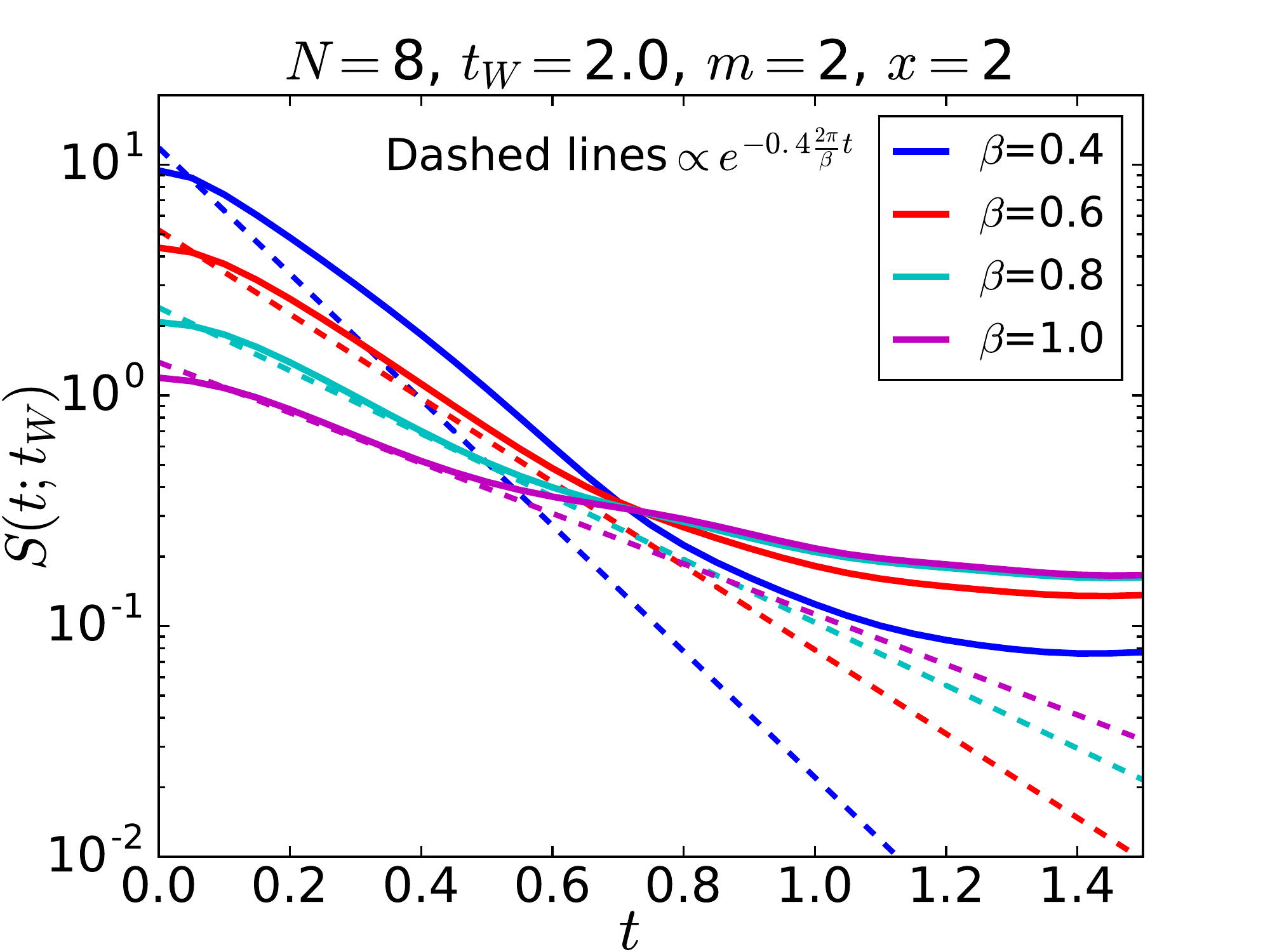}
  \includegraphics[width = 8cm]{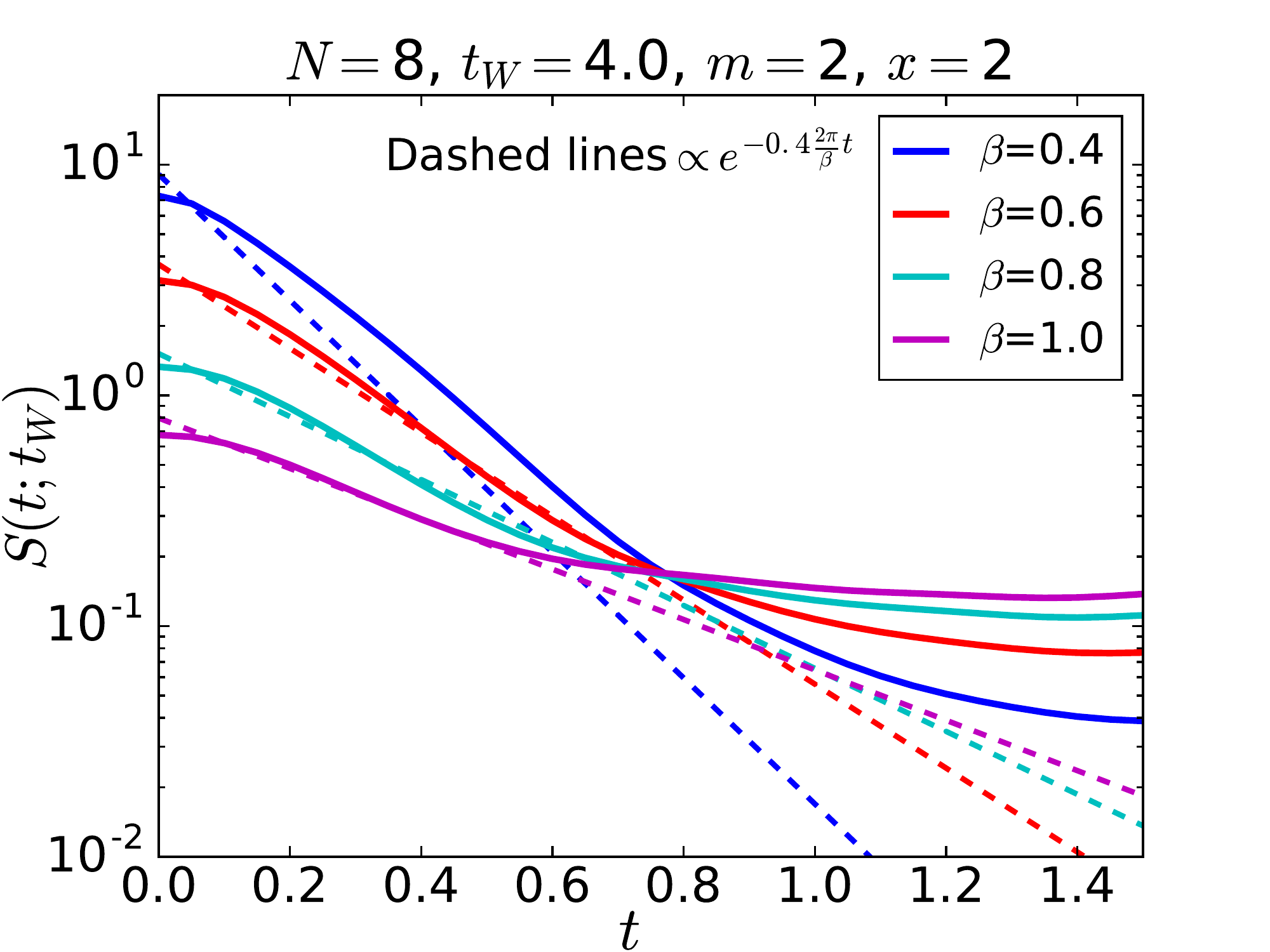}
 \caption{ Relative entropy $S(t; t_W)$ in the non-integrable model for $t_W=2$ (left) and $t_W=4.0$ (right).
  Dashed lines are proportional to $\exp( -0.4 \frac{2\pi}{\beta} t )$.  
 }
 \label{NI_t_dependence}
\end{figure*}

Numerical results in the non-integrable case ($g=-1.05, h=1$ in the model~\eqref{eq:SpinModel})
are presented in Fig.~\ref{NI_t_dependence}. 
As expected, the relative entropy decays exponentially at first and its rate is proportional to the temperature $\beta^{-1}$, which is qualitatively similar to what is predicted by conformal field theory considerations.
The initial exponential decay of the relative entropy is well explained by $\exp( -0.4 \frac{2\pi}{\beta} t )$
for $\beta \gtrsim 0.6$. For $\beta=0.4$ the exponent seems to be smaller. A possible reason for this is that since the butterfly velocity is $v_B = 2.5$ in this specific model~\cite{Hosur:2015ylk}, the thermal cycle is of one lattice size $v_B\beta \approx 1$ meaning that we have reached the cutoff temperature.

\begin{figure*}
  \includegraphics[width = 8cm]{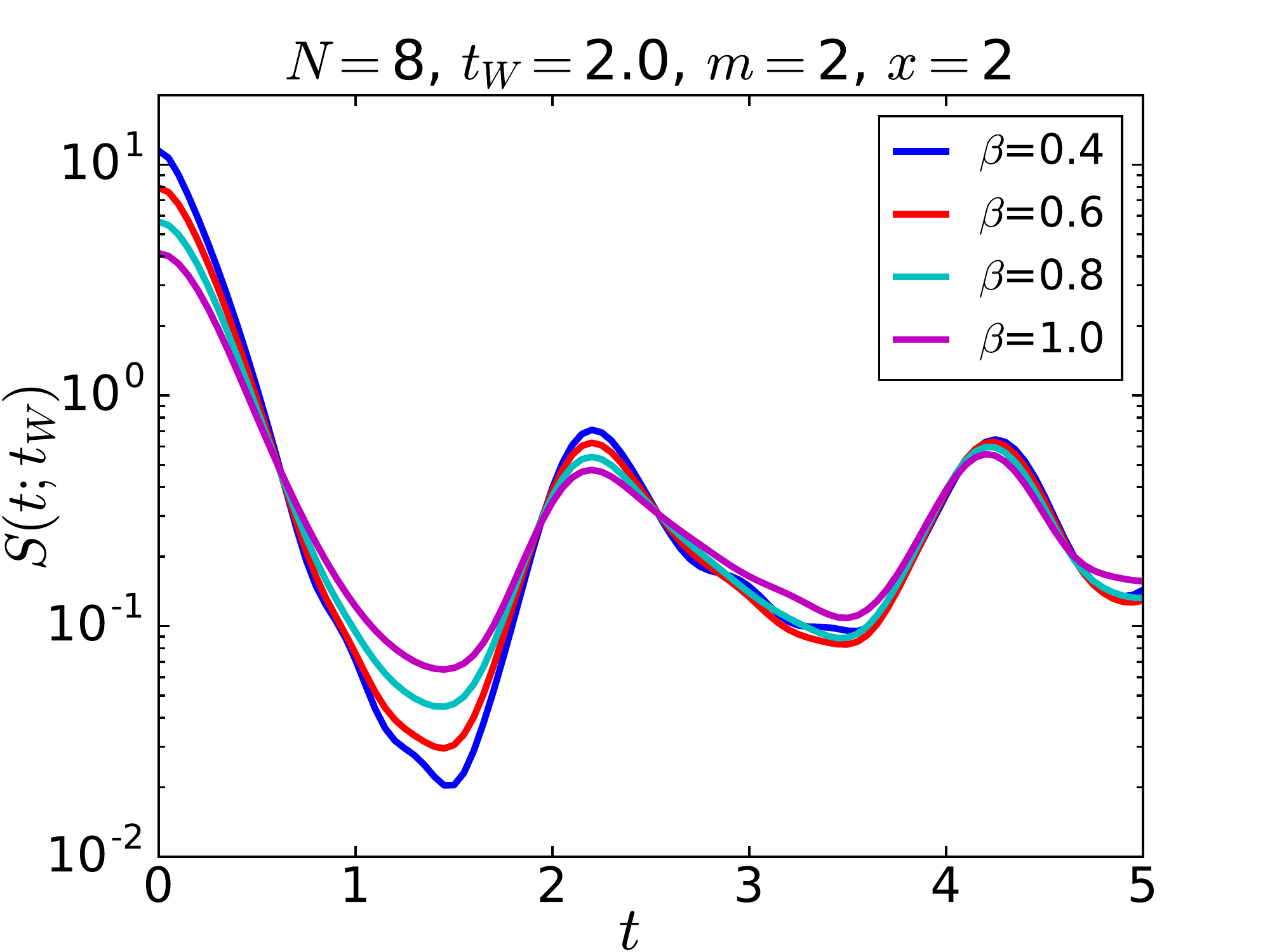}
  \includegraphics[width = 8cm]{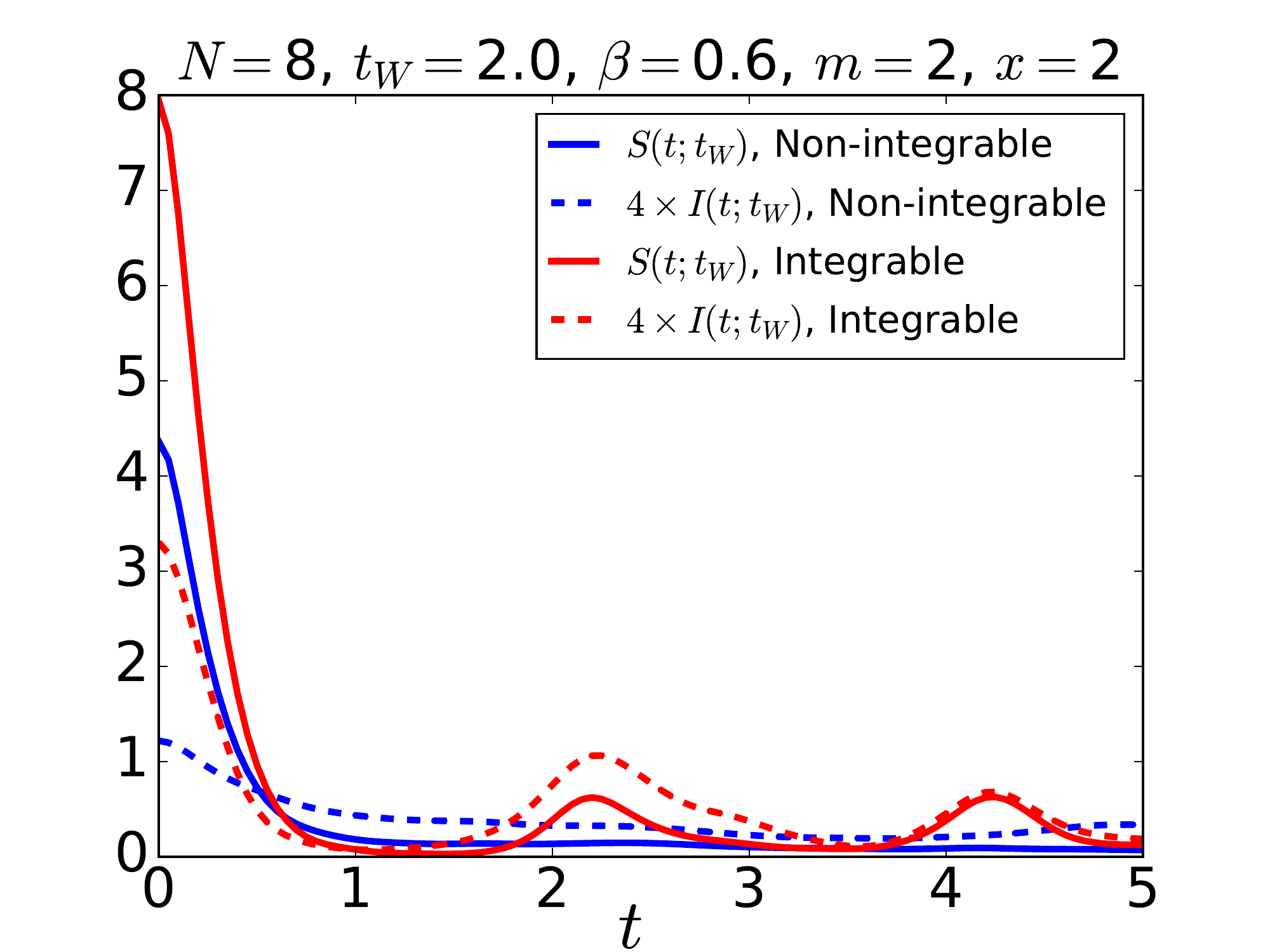}
 \caption{ (left) Relative entropy $S(t; t_W)$ in the integrable model for $t_W=2$ and various inverse-temperature $\beta$.
 (right) Comparison between the non-integrable model and integrable model for $t_W=2$ and $\beta=0.6$.
 }
 \label{IG_t_dependence}
\end{figure*}

The left panel of Fig.~\ref{IG_t_dependence} is the result for the integrable case ($g=1, h=0$).
As in the non-integrable case, the relative entropy decays exponentially at first (up to $t \sim 1.5$).
However, there is a revival of the relative entropy after the initial decay.
In the right panel of Fig.~\ref{IG_t_dependence}, we compare the results of the non-integrable model 
and the integrable model for $\beta=0.6$.
The relative strength of the revival is by far larger in the integrable case than in the non-integrable case.
We also note that the mutual information $I(t; t_W)$ is in phase with the relative entropy and exhibits the revival.
The revivals of the relative entropy and the mutual information are
clear manifestations of the integrability of the model.

\subsection{$t_W$-dependence}
Next we fix $t$ and investigate $t_W$-dependence of $S(t; t_W)$.
Again we take the size of each subsystem to be $m=2$ and the position of perturbation to be $x=2$.
In order to avoid numerical error in the relative entropy $S(t; t_W)$ due to very small eigenvalues ($\sim$ machine precision) of the density matrix, we set $t=0.1$ rather than $t=0$.

\begin{figure*}
  \includegraphics[width = 8cm]{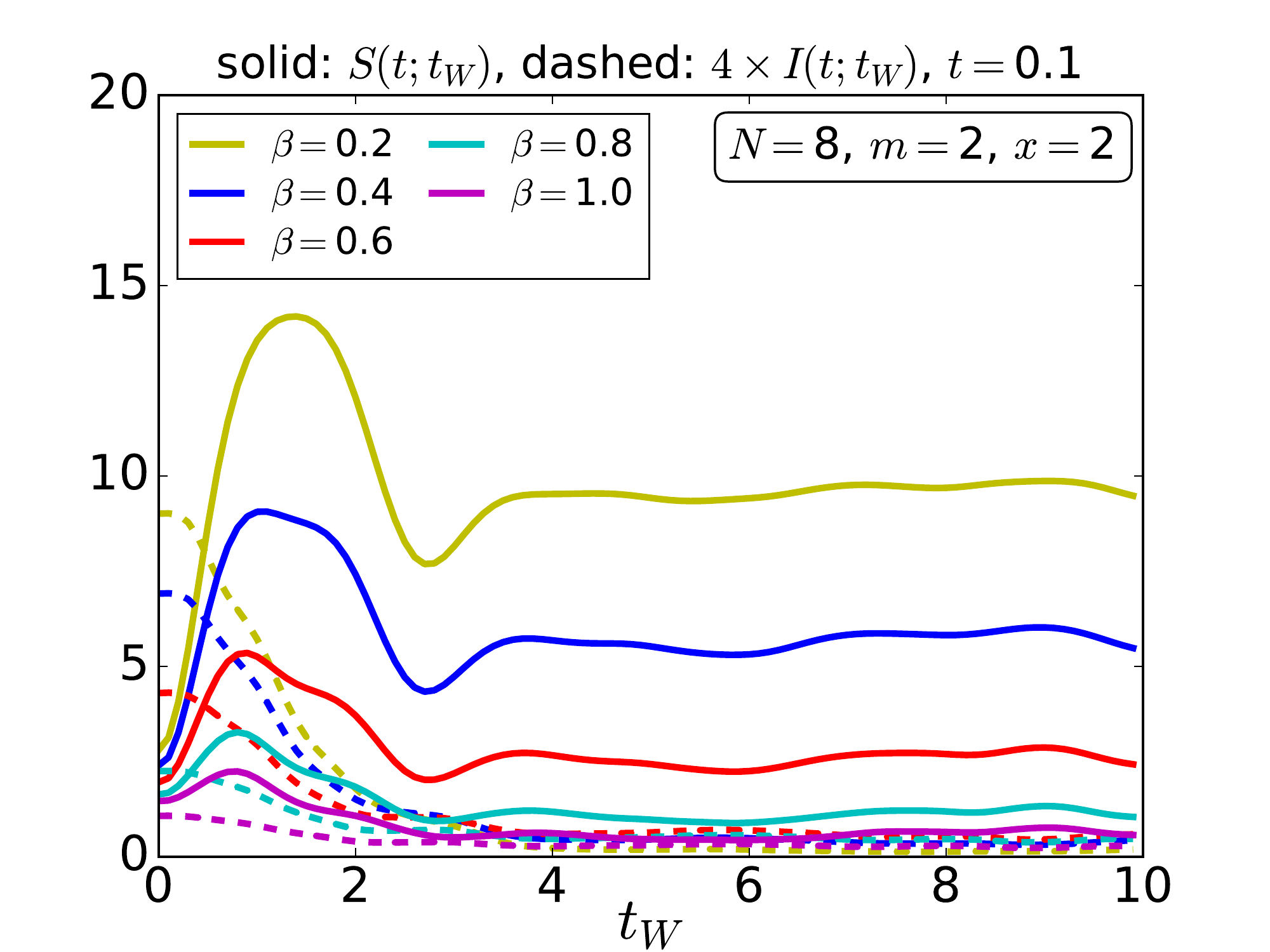}
  \includegraphics[width = 8cm]{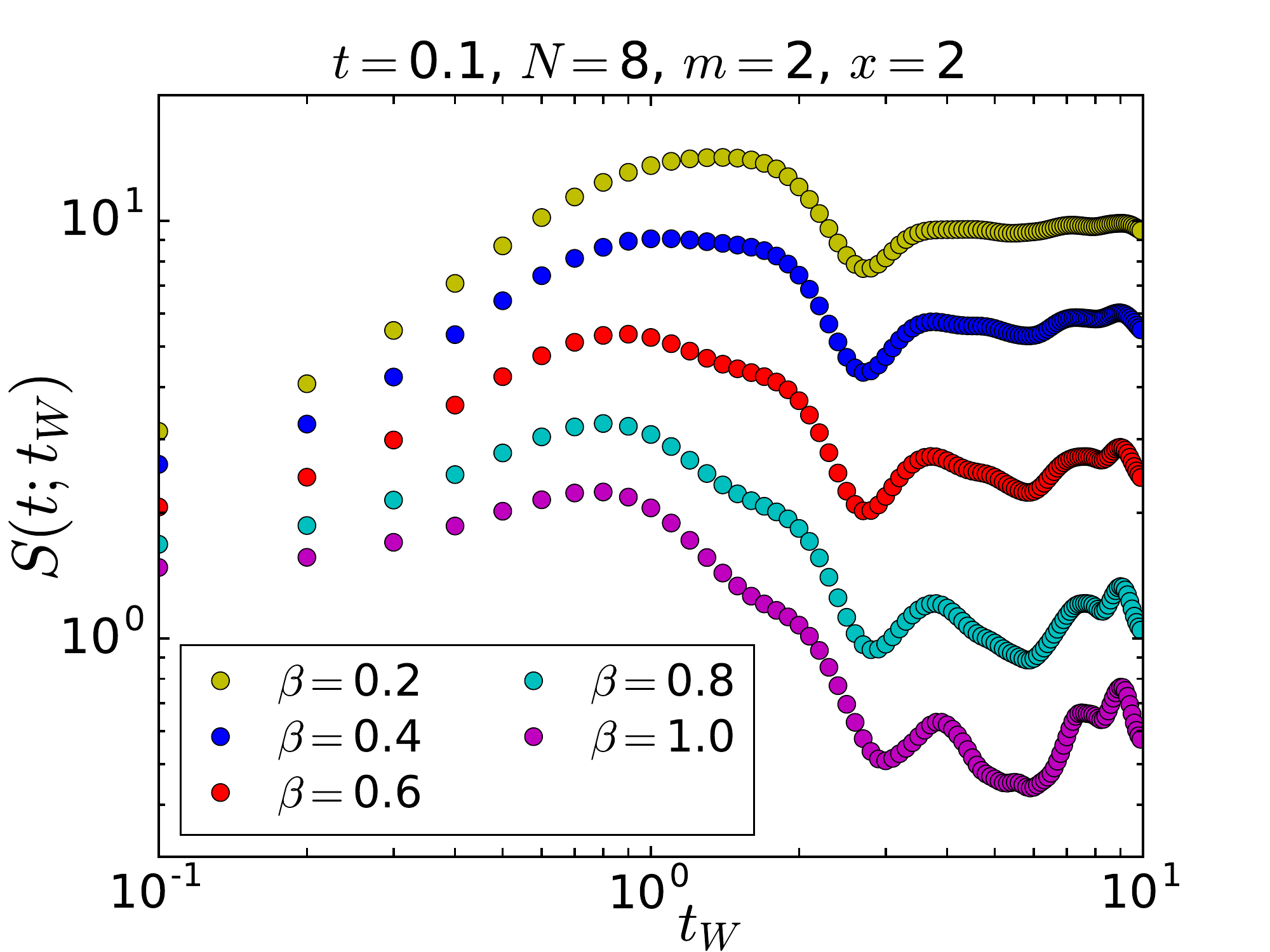}
 \caption{ $t_W$-dependence of the relative entropy $S(t; t_W)$ and the mutual information $I(t; t_W)$
 in the non-integrable model at $t=0.1$.
 Left panel is in a linear scale and right panel is in a log-log scale.
 }
 \label{NI_tw_dependence}
\end{figure*}

Figure~\ref{NI_tw_dependence} is a result for the non-integrable model ($g = -1.05, h = 0.5$ in~\eqref{eq:SpinModel}) at various inverse-temperature $\beta$.
$S(t; t_W)$ grows with $t_W$ at first but it starts to decay to some stationary values.
The inverse temperature $\beta$ affects the rate of the growth of the relative entropy, 
but dose not affect the time scale at which it becomes stationary.
We also observe that the time when $S(t; t_W)$ becomes stationary coincides with the time when the mutual information $I(t; t_W)$ decays to zero.
Moreover, the log-log plot (right panel of Fig.~\ref{NI_tw_dependence})
indicates that the initial growth of $S(t; t_W)$ in $t_W$ obeys a power-law,
although the holographic calculation predicts an exponential growth.
It would be interesting to study this algebraic growth in a spin chain from the viewpoint of the field theory. 

\begin{figure*}
  \includegraphics[width = 8cm]{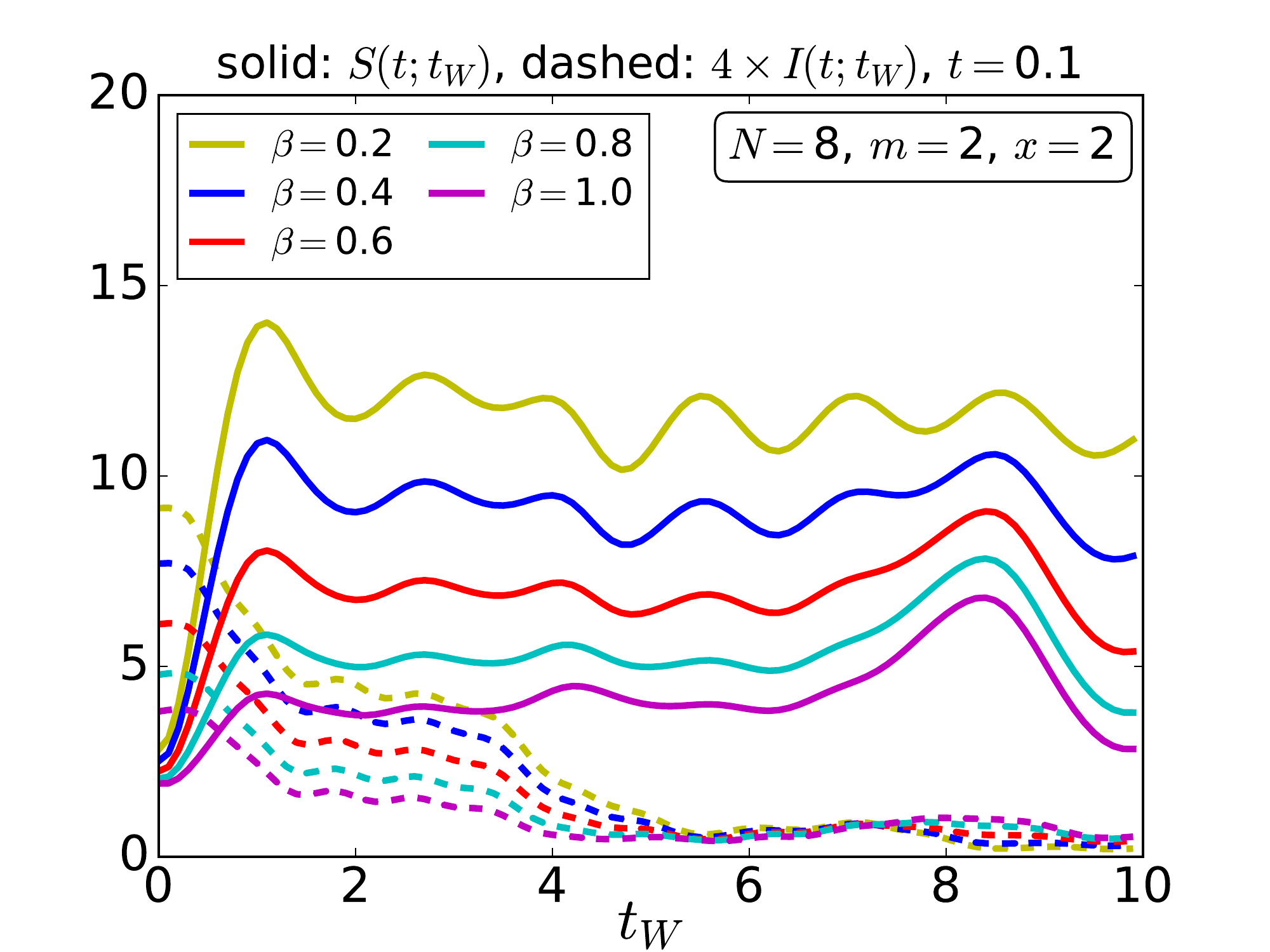}
  \includegraphics[width = 8cm]{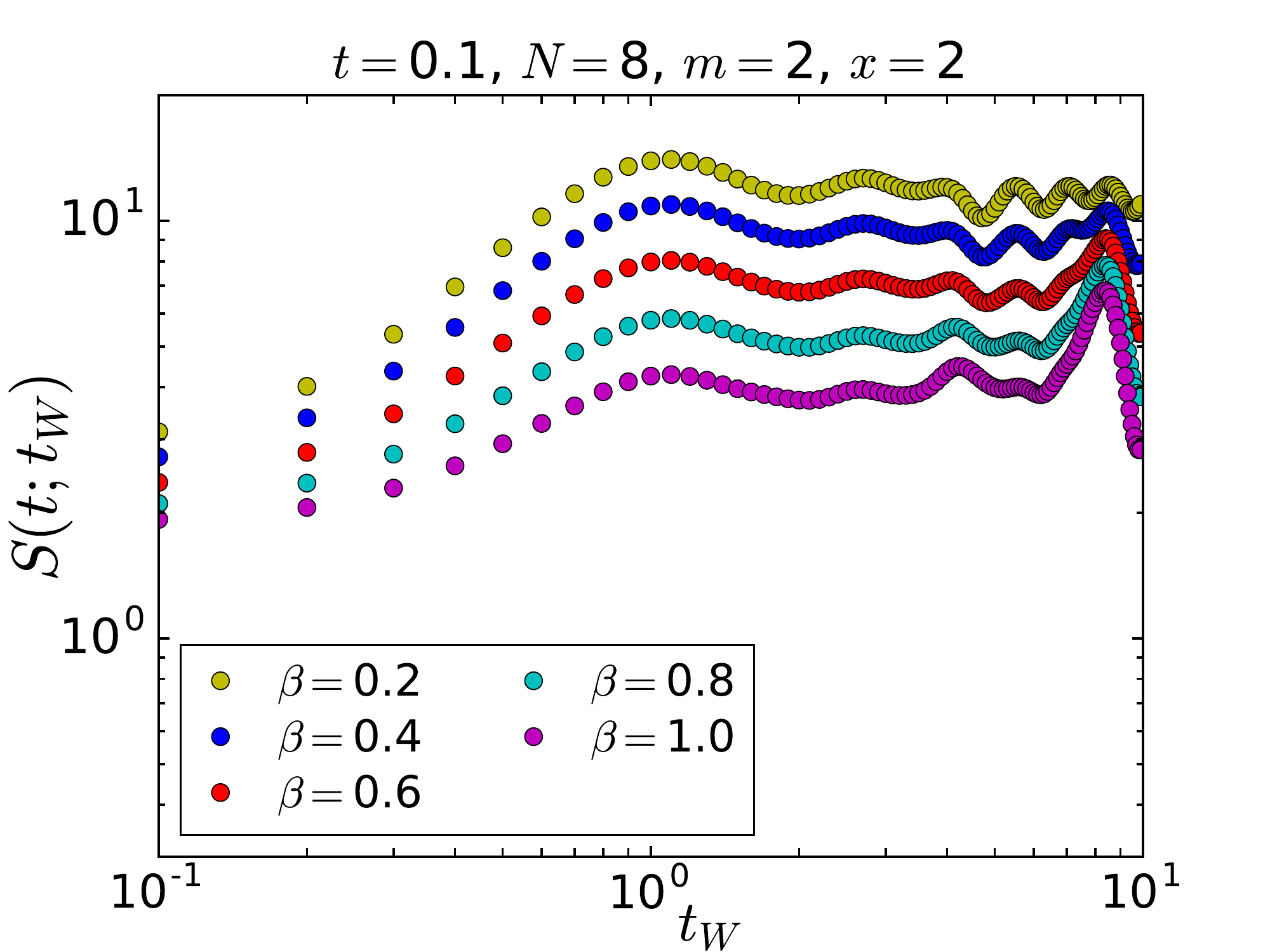}
 \caption{ $t_W$-dependence of the relative entropy $S(t; t_W)$ and the mutual information $I(t; t_W)$
 in the integrable model at $t=0.1$.
 Left panel is in a linear scale and right panel is in a log-log scale.
 }
 \label{IG_tw_dependence}
\end{figure*}

As for the integrable model, the qualitative behaviors of the relative entropy and the mutual information are almost the same as for the non-integrable model (Fig.~\ref{IG_tw_dependence}). 
Again, we observe an initial algebraic growth of the relative entropy and its saturation
in $t_W$.
However, the relative entropy shows a long-lived oscillations and a possible revival (around $t_W = 8.5$),
which results from the integrablity of the system.

\section{Discussions} 

We have considered the time evolution of relative entropy between the reduced density matrices of the thermofield double state and its perturbations. We have argued that the behavior of the relative entropy is in accord with the chaotic nature of the system. 
There are several things that could be investigated.  

\begin{itemize} 
\item Is there any bound on the decay rate of the relative entropy?  We now know that the growth rate of both entanglement entropy and OTO correlators (ie, butterfly velocity $v_{B}$ and entaglement velocity $v_{E}$ respectively) are bounded by the corresponding thermal quantities \cite{Maldacena:2015waa, Hartman:2015apr}. It seems natural to anticipate that our two sided relative entropy is bounded as well. We have discussed a possible lower bound \eqref{eq:decaylowerbound} on the decay rate assuming the chaos bound continues well for $n\geq 1$. It is not clear however if this assumption can be justified. Also, one might anticipate the existence of an upper bound on the decay rate.
\item We have seen that the relative entropy grows exponentially with the insertion time of the perturbing operator, and that this growth is fairly universal with an exponent $2\pi/\beta$. It would be interesting to see if this can be used to understand the chaos bound of \cite{Maldacena:2015waa} from an information theoretic point of view.
\item It might be interesting to ask what is the holographic relative entropy doing after the scrambling time. An eternal decay such as in \eqref{eq:generalform} might be in conflict with unitarity because it would mean that the density matrices become identical, though we do not have a tight argument that this is not possible as for the thermal two point function \cite{Maldacena:2001kr} or the spectral form factor \cite{Papadodimas:2015xma}.
\item Higher dimensional generalization of the result. One can in principle calculate the holographic entanglement entropy in the presence of a shock wave in higher dimensions, at least when the subsystem is the half of the total spatial manifold \cite{Shenker:2013pqa}. Once the entanglement entropy is known, one can obtain the relative entropy by removing the first law term. 
\item There are other black holes with a long wormhole throat in the horizon interior. For example  \cite{Bak:2007jm}.  The time evolution of entanglement entropy was discussed in \cite{Ugajin:2014nca}. It would be 
interesting to generalize our analysis to this case. 
\end{itemize}

\subsection*{Acknowledgment}

We thank discussions to Pawel Caputa, Nima Lashkari, Aitor Lewkowycz, Henry Maxfield, Masahiro Nozaki and Onkar Parrikar.

Y.O.N. was supported by Advanced Leading Graduate Course for Photon Science (ALPS) of the Japan Society for the Promotion of Science (JSPS) and by JSPS KAKENHI Grants No. JP16J01135.
The work of G.S. was supported in part by a grant from the Simons Foundation (\#385592, Vijay Balasubramanian) through the It From Qubit Simons Collaboration, by the Belgian Federal Science Policy Office through the Interuniversity Attraction Pole P7/37, by FWO-Vlaanderen through projects G020714N and G044016N, and by Vrije Universiteit Brussel through the Strategic Research Program ``High-Energy Physics''.

\appendix

\section{The out of causal contact case for localized perturbations} 
 \label{app:outofcausal}
 
The only case we can always solve analytically is when the insertion point is in the domain of dependence of the traced out region on Fig \ref{fig:1}, i.e.  $-x>t_p-t>x$.  Let us summarize the result in this case. We just briefly adapt the calculation in \cite{Sarosi:2016oks,Sarosi:2016atx} to the present situation without going into too much details. 

As mentioned earlier, as $\epsilon \rightarrow 0$ we are in the OPE limit
\beq
z_{1;k} \rightarrow z_{2;k}, \quad {\bar z}_{1;k} \rightarrow {\bar z}_{2;k},
\eeq
and we can obtain the result in a similar way as done in a small subsystem size limit. We will assume $x>0$. We have the OPE
\bea
&V(z_{1;k},\bar { z}_{1;k})V( z_{2;k},\bar { z}_{2;k}) \\&=\langle VV\rangle \Big(1+C^O_{VV} (e^{2\pi i k/n}2\pi i\epsilon g_n)^{h_O} (-e^{-2\pi i k/n}2\pi i\epsilon \bar g_n)^{\bar h_O} O(e^{2\pi i k/n} u_n,e^{-2\pi i k/n} \bar u_n) + \cdots\Big),
\eea
where $O$ is the lightest primary in the OPE and we introduced the notation
\bea
u_n &=e^{-\frac{2\pi}{\beta n}t}\left( \frac{\sinh{\frac{\pi (w_{*}+t)}{\beta}}}{\cosh{\frac{\pi( w_{*}-t)}{\beta}} } \right)^{\frac{1}{n}},\\
g_n &=u_n \frac{1}{ \beta n} \left(\coth \frac{\pi(w_*+t)}{\beta} -\tanh \frac{\pi(w_*-t)}{\beta}\right),\\
\eea
and the barred counterparts have $w_* \rightarrow \bar w_*$ with $w_*=x-t_p$, $\bar w_*=x+t_p$ and $t\rightarrow -t$. We need to do the analytic continuation for
\beq \sum_{k \neq l=0}^{n-1} \langle O(e^{2\pi i k/n} u_n)O(e^{2\pi i l/n} u_n) \rangle.
\eeq
However, we may scale out the common $u_n$ factors using global conformal invariance and write
\beq
 \frac{1}{u_n^{h_O}\bar u_n^{\bar h_O}}\sum_{k \neq l=0}^{n-1} \langle O(e^{2\pi i k/n} )O(e^{2\pi i l/n} ) \rangle,
\eeq
which is the same correlator that needs analytic continuation in the small subsystem size limit for global states, see \cite{Sarosi:2016oks,Sarosi:2016atx} for the details. The replica relative entropy then reads
\beq
S_n(\rho_V||\rho_W) = \frac{f(\Delta,n)}{1-n} \left( \frac{n}{2} (C^O_{VV})^2-C^O_{WW}C^O_{VV}-\frac{n-2}{2} (C^O_{WW})^2\right) \left[2\pi i \frac{\epsilon g_n}{\sqrt{u_n}} \right]^{2h_O} \left[-2\pi i \frac{\epsilon \bar g_n}{\sqrt{\bar u_n}} \right]^{2 \bar h_O},
\eeq
with
\beq
f(\Delta,n)\sim \sum_{k \neq l=0}^{n-1} \langle O(e^{2\pi i k/n} )O(e^{2\pi i l/n} ) \rangle.
\eeq
Taking $n\rightarrow 1$ leads to 
\bea
S(\rho_V||\rho_W)& \sim  (C^O_{VV}-C^O_{WW})^2  \\ &\times  \left[\epsilon^2  \frac{e^{\frac{2\pi t}{\beta}}\cosh^2\frac{2\pi}{\beta}t}{2\beta^2 \cosh\frac{2\pi}{\beta}(x-t_p-t)\sinh^3\frac{2\pi}{\beta}(x-t_p+t)}\right]^{h_O} \\ &\times \left[ \epsilon^2  \frac{e^{-\frac{2\pi t}{\beta}}\cosh^2\frac{2\pi}{\beta}t}{2\beta^2 \cosh\frac{2\pi}{\beta}(x+t_p+t)\sinh^3\frac{2\pi}{\beta}(x+t_p-t)}\right]^{\bar h_O},
\eea
where we have neglected some $\Gamma$ function factors depending on the dimensions of $O$ for simplicity, they are the same as in \cite{Sarosi:2016oks,Sarosi:2016atx}. There is an expected singularity as the operator insertion approaches the light cones on Fig. \ref{fig:1} (reminder: the above formula is valid in the left wedge), so this formula is valid as long as $|\beta\epsilon^2(x+t_p-t)^{-3}| \ll 1$ and $|\beta\epsilon(x-t_p+t)^{-3}|\ll 1$. Higher orders in $\epsilon$ in the left wedge are systematicaly obtainable using modular perturbation theory techniques \cite{Sarosi:2017rsq}.

\section{Relative entropy between two perturbed states} 
\label{sec:distinctstates}

Here we give a generalization of the large $c$ vacuum block result \eqref{eq:generalform} for the relative entropy to the case when both states have operator insertions with different dimensions, i.e. we study the relative entropy $S(\rho_{V}|| \rho_{W})$ between two states of the form 
\be
V(t_{p} -i \epsilon,x) |TFD\ra,  \quad W(t_{p} -i \epsilon,x) |TFD\ra.
\ee  
The situation we tackle is when $V$ is an arbitrary primary, while $W$ is some uniformized operator creating a conformal transformation of the thermofield double. In holography, any state dual to a Ba\~nados geometry can be treated this way.
Now
\be 
S(\rho_{V}|| \rho_{W}) = {\rm tr} \left[K_{W} \delta \rho \right] - \left[ S(\rho_{V}) -S(\rho_{W}) \right], \label{eq:relshock}
\ee
where $\delta \rho =\rho_{V}-\rho_{W}$. The second term is just the difference of entanglement entropies  $S(h_{V})-S(h_{W})$ which we can easily obtain from \eqref{eq:topwedgeEE}. Because of the first law, to linear order in $h_V-h_W$, the first term of the relative entropy (\ref{eq:relshock}) is given by the derivative of the entanglement entropy \eqref{eq:topwedgeEE} as a function of $h$
\begin{align}
 {\rm tr} \left[K_{W} \delta \rho \right] &\approx S'(h_{W}) (h_{V} -h_{W}) \nonumber  \\ 
&=\f{c}{6} \f{(h_{V}-h_{W})F(t)}{1+h_{W}F(t)}, \quad F(t)=\frac{ \pi}{3c}  \frac{\sinh \frac{2\pi t}{\beta}+\sinh \frac{2\pi (x-t_p)}{\beta}}{\sin \frac{2\pi \epsilon}{\beta}\cosh\frac{2\pi t}{\beta}}.
\label{eq:first}
\end{align}
The above formula should be valid to linear order in $h_V-h_W$ and in the bottom wedge of Fig. \ref{fig:1} and it is easy to obtain the analogous formula for the top wedge. Further progress can be made by restricting to the case when the state $|W\rangle$ is a conformal transformation of the TFD state, since in this case it has a local modular Hamiltonian that is an integral of the stress tensor and therefore the linear order in $h_V-h_W$ expression is exact. The application of this ``first law trick" also relies on the assumption that we can continuously turn off the perturbation in $S(\rho_W||\rho_{TFD})$. We do not expect this to be true in the right wedge (causal diamond of the subsystem) since in this case the first law is already violated in $S(\rho_W||\rho_{TFD})$ as $h_W\rightarrow 0$. This is because the modular Hamiltonian expectation value \eqref{eq:rightwedge} is $\sim \epsilon^{-1}$ while the entanglement entropy vanishes as $\epsilon \rightarrow 0$. The distinguishing feature of the result \eqref{eq:generalform} allowing this trick to work is that it only depends on the energy $E_W\sim h_W/\sin \epsilon$ of the state and not on its coupling to other operators.\footnote{One can also play this trick for thermal states on the line, where the modular Hamiltonian is known and it can be easily verified that the trick works, see \cite{Sarosi:2016atx}.  }

In the bottom wedge for such states we therefore have
\beq
S(\rho_V||\rho_W) \approx \f{c}{6} \f{(h_{V}-h_{W})F(t)}{1+h_{W}F(t)}-\frac{c}{6} \log \left( \frac{1+h_{V}F(t)}{1+h_{W}F(t)}\right) 
\eeq
Notice that in the regime $F(t)\ll h_W^{-1}$, this still shows an exponential decay
\beq
S(\rho_V||\rho_W) \sim \frac{c}{12}(h_V-h_W)^2 F(t)^2,
\eeq
with exponent $\frac{4\pi}{\beta}$. On the other hand, for early times ($F\gg h_V^{-1},h_W^{-1}$) the exponential decay is absent from the modular Hamiltonian part, whenever $h_W\neq 0$.

\section{Relative entropy of two disjoint intervals}  
\label{app:disjoint}
In this section we generalize the calculation of the  relative entropy between the states $| TFD \ra, | \Psi_{S} \ra$ to include the effect of a finite subsystem size. We consider the case where the subsystem is the union of two disjoint intervals, one is in the left CFT, and the other is in the right. 

We take the disjoint union of two intervals $ A \cup B$, whose end points are  
\be 
P_{1} : \left (-\f{x}{2}, t_{L} =-t\right) , \quad P_{2} : \left (\f{x}{2},  t_{L} =-t \right), \quad P_{3} :\left (-\f{x}{2}, t_{R} =t\right), \quad   P_{4} :\left (\f{x}{2}, t_{R} =t\right) \label{eq:ends}
\ee
Let $\gamma_{ij} $ be the bulk geodesics connecting $P_{i} $ and $P_{j}$ and $L_{ij} $ be the length of the curve $\gamma_{ij}$.  The holographic entanglement entropy is given by 
\be 
S_{A\cup B} = \f{1}{4G_{N}} {\rm min} \left[ L_{12} +L_{34}, \; L_{13}+ L_{24}  \right] .
\ee
Hereafter we denote 
\be 
 \quad S_{c} \equiv\f{ L_{13} +L_{24}}{4G_{N}},\quad  S_{d} \equiv\f{ L_{12} +L_{34}}{4G_{N}}. 
\ee

\subsection{Holographic entanglement entropy}

\subsubsection*{BTZ black hole}  

In this case \cite{Hartman:2013qma}
\be 
S_{c}  = \f{2c}{3}\log \f{r_{\infty}}{R} + \f{2c}{3}\log \left[ \cosh \f{2\pi t}{\beta}\right] , \quad  S_{d}  = \f{2c}{3}\log \f{r_{\infty}}{R} + \f{2c}{3}\log \left[ \sinh \f{\pi x}{\beta}\right] 
\ee
where $r_{\infty}$ denotes the UV cut off.    In the high temperature limit $x \gg \beta$, the entanglement entropy is given by   
\be 
S_{A\cup B}(\rho_{TFD}) = \begin{cases}
   S_{c} &  t \leq \f{x}{2} \\
   S_{d}& t > \f{x}{2}
  \end{cases}. 
\ee

\subsubsection*{Shockwave wave geometry} 

 In the shock wave limit (\ref{eq:midya})
$S_{c}$, $S_{d}$ are  given by  \cite{Shenker:2013pqa}
\be 
S_{c} = \f{c}{3}\log \f{r_{\infty}}{R} + \f{c}{3}\log \left[ \cosh \f{2\pi t}{\beta} + \f{\alpha}{2} \right], \quad  S_{d} = \f{2c}{3}\log \f{r_{\infty}}{R} + \f{2c}{3}\log \left[ \sinh \f{\pi x}{\beta}\right] 
\ee
When $\alpha$ is smaller than the critical value  $\alpha_{*}$ 
\be
\alpha \leq \alpha_{*}, \qquad  1+\f{\alpha_{*}}{2} = \sinh \f{\pi x}{\beta} \label{eq:crital},
\ee
the holographic entanglement entropy is given by 
\be 
S_{A\cup B} = \begin{cases}
  \f{c}{3}\log \f{r_{\infty}}{R} + \f{c}{3}\log \left[ \cosh \f{2\pi t}{\beta} + \f{\alpha}{2} \right], &  t \leq t_{*} \\[+10pt]
   \f{2c}{3}\log \f{r_{\infty}}{R} + \f{2c}{3}\log \left[ \sinh \f{\pi x}{\beta}\right] ,& t > t_{*}
  \end{cases}, 
\quad \cosh \f{2\pi t_{*}}{\beta} + \f{\alpha}{2}  =  \sinh \f{\pi x}{\beta} \label{eq:small}
\ee
However, when $\alpha >\alpha_{*} $ there is no phase transition analogous to the BTZ case in the entanglement entropy, as $S_{d} $ always gives dominant contribution.   
\be
S_{A\cup B} =\f{2c}{3}\log \f{r_{\infty}}{R} + \f{2c}{3}\log \left[ \sinh \f{\pi x}{\beta}\right]  \label{eq:large}.
\ee

\subsection{Modular Hamiltonian of the TFD state for two disjoint intervals}
\label{section:modH}

\subsubsection*{Vacuum modular Hamiltonian for two disjoint intervals in the large $c$ limit}  

It is hard to analytically obtain the modular Hamiltonian for two disjoint intervals $A\cup B$ even for the vacuum state. However, in the large central charge limit this modular Hamiltonian must have a simple expression because of its relation to the bulk area operator in the dual gravity theory \cite{Jafferis:2015del}
\be
K = \f{A}{4G_{N}} +o(1).
\ee
If we we denote $w_{i}$ by the holomporphic coordinate of the end point $P_{i}$ of the subsystem $A \cup B$, then the holomorphic part of the modular Hamiltonian $K^{(0)}_{A\cup B}$ is
\be 
K^{(0)}_{A\cup B}= \begin{cases}
   K^{(0)}_{w_{1},w_{2}} + K^{(0)}_{w_{3},w_{4}}&  w \leq 1 \\[+10pt]
  K^{(0)}_{w_{1},w_{3}} + K^{(0)}_{w_{2},w_{4}} & w>1
  \end{cases} \quad  \quad w= \f{(w_{1}-w_{2})(w_{3}-w_{4})}{(w_{1}-w_{3})(w_{2}-w_{4})}.  \label{eq:moddis}
\ee

\be
K^{(0)}_{w_{1},w_{2}} = \int^{w_{2}}_{w_{1}} \f{(w_{2}-w)(w-w_{1})}{w_{2}-w_{1}} T_{ww}(w) dw
\ee
where $T_{ww}(w)$ is the holomorphic part of stress tensor. 
We also have similar expression for the anti holomorphic part.

\subsubsection*{Modular Hamiltonian of the TFD state} 

The modular Hamiltonian of a TFD state is then given by conformal mapping of the result (\ref{eq:moddis})

\be 
K_{T} = \begin{cases}
K_{(c)} =K_{13}+ K_{24}+c.c&  t \leq \f{x}{2} \\[+10pt]  
K_{(d)}=K_{12}+K_{34}+c.c & t \geq \f{x}{2} 
 \end{cases} ,
\label{eq:modhth}
\ee
with 
\be 
K_{12} = \f{\beta}{2\pi \sinh \f{\pi x}{\beta}} \int^{\f{x}{2}}_{-\f{x}{2}} dy \left( \cosh \f{\pi x}{\beta} - \cosh \f{2\pi y}{\beta}\right) T_{zz} (-t+ i\f{\beta}{2},y),
\ee

\be 
K_{34} = \f{\beta}{2\pi \sinh \f{\pi x}{\beta}} \int^{\f{x}{2}}_{-\f{x}{2}} dy \left( \cosh \f{\pi x}{\beta} - \cosh \f{2\pi y}{\beta}\right) T_{zz} (t,y),
\ee

\be 
K_{13}= \f{\beta}{2\pi \cosh \f{2\pi t}{\beta}} \int^{t}_{-t + \f{i\beta}{2}}  dy\left( \sinh \f{2\pi t}{\beta} -\sinh \f{2\pi y}{\beta} \right) T_{zz} (y, -\f{x}{2}) ,
\ee 

\be 
K_{24}= \f{\beta}{2\pi \cosh \f{2\pi t}{\beta}} \int^{t}_{-t + \f{i\beta}{2}} dy \left( \sinh \f{2\pi t}{\beta} -\sinh \f{2\pi y}{\beta} \right) T_{zz} (y, \f{x}{2}) .
\ee
Let us evaluate expectation values of these operators on the  state $\rho_{S}$. For $K_{12}, K_{34}$, these values are vanishing in the Vaidya limit  (\ref{eq:midya})
\be 
{\rm tr} \rho_{S} \;K_{12} ={\rm tr} \rho_{S} \;K_{34}   =0, \rightarrow  {\rm tr} \rho_{S} \;  K_{(d)}=0.
\ee 
For $K_{(c)}=K_{13},+K_{24}+c.c$, since 
\be 
{\rm tr} \left[ T_{00} (t_{L},x) \rho_{S} \right] = \begin{cases}
\f{E}{2\pi}&  t_{L} \leq t_{W} \\
0& t_{L} \geq t_{W}
 \end{cases}, 
\quad {\rm tr} \left[T_{00} (t_{R},x) \;\rho_{S} \right]=0,
\ee
we have 
\begin{align} 
{\rm tr}  K_{(c)}  \rho_{S}  &= {\rm tr} \left[ (K_{13},+K_{24}+c.c )\;\rho_{S} \right] \\
 &= \f{2\beta E}{2 \pi \cosh \f{2\pi t}{\beta} }  \int^{t_{W} +\f{i\beta}{2}}_{-t+\f{i\beta}{2}}\left[\sinh \f{2\pi t}{\beta} -\sinh \f{2\pi y}{\beta} \right] dy  \\[+10pt]
&\rightarrow  \left(\f{\beta}{2\pi} \right)^2 \f{4M \alpha}{ \cosh \f{2\pi t}{\beta}},
\end{align}
in the Vaidya limit (\ref{eq:midya}).  By using the relation between the mass and temperature of the black hole
\be 
M= \f{c}{12} \left( \f{2\pi}{\beta}\right)^2,
\ee
We have 
\be 
{\rm tr}  K_{(c)} \rho_{S}= \f{c}{3} \f{\alpha}{ \cosh \f{2\pi t}{\beta}}.
\ee
This expression is plausible since it obeys the first law relation 
\be 
\delta S_{c} = {\rm tr}  K_{(c)}  \; \rho_{S} + O(\alpha^2).
\ee

\subsection{Relative entropy}

By combining these results  we obtain the expressions of the relative entropy.  
When $\alpha <\alpha_{*}$, from  (\ref{eq:small}) and (\ref{eq:modhth}) we obtain
\begin{align}
S( \rho_{S} || \rho_{TFD} ) =\begin{cases} 
\f{c}{3} \f{\alpha}{ \cosh \f{2\pi t}{\beta}}  -  \f{2c}{3}\left(\log \left[ \cosh \f{2\pi t}{\beta}+\f{\alpha}{2} \right]- \log  \cosh \f{2\pi t}{\beta}\right) &  t \leq t_{*}\\[+10pt]
\f{c}{3} \f{\alpha}{ \cosh \f{2\pi t}{\beta}}  -  \f{2c}{3}\left(\log  \sinh \f{\pi x}{\beta}- \log  \cosh \f{2\pi t}{\beta}\right) &   t_{*} \leq  t \leq \f{x}{2}\\[+10pt]
0 &    \f{x}{2} \leq t
\end{cases} .
\end{align}
For $t<t_*$ the relative entropy is of the general form \eqref{eq:generalform} that we observed for infinite subsystems. However, at time $t_*$, depending on the size of the subsystem via \eqref{eq:small}, the form of the decay changes and the relative entropy reacher zero at time $t=\frac{x}{2}$. This shows that the decay is controlled by the size of the subsystem whenever this is smaller than the scrambling time $\beta \log c$.

When $\alpha >\alpha_{*}$ 
 , from (\ref{eq:large})  and (\ref{eq:modhth}) we have
\begin{align} 
S( \rho_{S} || \rho_{TFD} ) =\begin{cases} 
\f{c}{3} \f{\alpha}{ \cosh \f{2\pi t}{\beta}}  -  \f{2c}{3}\left(\log  \sinh \f{\pi x}{\beta}- \log  \cosh \f{2\pi t}{\beta}\right) &  t \leq \f{x}{2}\\[+10pt]
0 &  t \geq \f{x}{2}
\end{cases} 
\end{align}
In the expression of the first line, the first term is of order $e^{\f{2\pi l}{\beta}}$ because of the critical value (\ref{eq:crital}), while the second term is of order  $\f{2\pi l}{\beta}$, therefore the fist term dominates. This means that the relative entropy is exponentially decaying in time, but we clearly see that the value of the relative entropy stays large until $t=\f{x}{2}$ . This is because the initial difference between two reduced density matrices $\rho_{S}(0)$ and $\rho_{TFD}(0)$ is too large for the system to scramble the quantum information of $\rho_{S}$ by $t=\f{x}{2}$. Indeed, $\rho_{S}$ has factorized form even at $t=0$, $\rho_{S}= \rho_{A} \otimes \rho_{B}$. This follows from the fact that the mutual information $I_{AB}$ of $\rho_{S}$ vanishes, and $I_{AB} (\rho_{S}) =S(\rho_{AB} ||\rho_{A} \otimes \rho_{B})$.  On the other hand it takes $t=x/2$ time for $\rho_{TFD}$ to get factorized, and this is the reason why the relative entropy stays large during the process. 

\bibliographystyle{utphys}
\bibliography{chaosandrelent}

\end{document}